\documentclass[12pt]{article}
\usepackage{amsmath}
\usepackage{color}
\usepackage[unicode,bookmarks,bookmarksopen,bookmarksopenlevel=2,colorlinks,linkcolor=blue,citecolor=green]{hyperref}

\input{tcilatex}
\begin{document}

\title{\textbf{Integrable hydrodynamic chains associated with Dorfman
Poisson brackets}}
\author{Maxim V. Pavlov \\
\\
Department of Mathematical Physics, Lebedev Physical Institute,\\
Russian Academy of Sciences, Moscow}
\date{}
\maketitle

\begin{abstract}
This paper is devoted to a description of integrable Hamiltonian
hydrodynamic chains associated with Dorfman Poisson brackets. Three main
classes of these hydrodynamic chains are selected. Generating functions of
conservation laws and commuting flows are found. Hierarchies of these
Hamiltonian hydrodynamic chains are extended on negative moments and
negative time variables. Corresponding three dimensional quasilinear
equations of the second order are presented.
\end{abstract}

\tableofcontents

\newpage\ \ \ \ \ \ \ \ \ \ \ \ \ \ \ \ \ \ \ \ \ \ \ \ \ \ \ \ \ \ \ \ \ \
\ \ \ \ \ \ \ \ \ \ \ \ \ \ \ \ \ \ \ \ \ \ \ \ \ \ \ \ \ \textit{\ \ \ \ \
\ \ \ \ \ \ \ \ \ \ \ On the occasion}

\textit{\ \ \ \ \ \ \ \ \ \ \ \ \ \ \ \ \ \ \ \ \ \ \ \ \ \ \ \ \ \ \ \ \ \
\ \ \ \ \ \ \ \ \ \ \ \ \ \ \ \ \ \ \ \ \ \ \ of Eugene Ferapontov's 50th
birthday}

\section{Introduction}

Recently (see \cite{MaksZyk}), integrable hydrodynamic chains written in the
conservative form%
\begin{equation}
\partial _{t}h_{k}=\partial _{x}f_{k}(h_{0},h_{1},...,h_{k+1}),\text{ \ }%
k=0,1,2,...  \label{e}
\end{equation}%
were completely described (this classification problem was established in 
\cite{MaksEps}, the integrability criterion based on the differential
geometric approach was suggested in \cite{FerMarsh}, where first two
expressions $f_{0}(h_{0},h_{1})$ and $f_{2}(h_{0},h_{1},h_{2})$ were
determined). Also, it was proved that these conservative hydrodynamic chains
can be written via special coordinates, i.e. the so called moments $%
A^{k}(h_{0},h_{1},...,h_{k})$ such that%
\begin{equation}
A_{t}^{k}=f_{1}A_{x}^{k+1}+f_{0}A_{x}^{k}+A^{k+1}(s_{0}A_{x}^{0}+s_{1}A_{x}^{1})+A^{k}(r_{0}A_{x}^{0}+r_{1}A_{x}^{1})
\label{lin}
\end{equation}%
\begin{equation*}
+k[A^{k+1}(w_{0}A_{x}^{0}+w_{1}A_{x}^{1}+w_{2}A_{x}^{2})+A^{k}(v_{0}A_{x}^{0}+v_{1}A_{x}^{1}+v_{2}A_{x}^{2})+A^{k-1}(u_{0}A_{x}^{0}+u_{1}A_{x}^{1}+u_{2}A_{x}^{2})],
\end{equation*}%
where coefficients $f_{i},s_{j},r_{k}$ depend on first two moments $A^{0}$
and $A^{1}$ only, while all other coefficients $w_{m},v_{n},u_{p}$ depend
just on first three moments $A^{0},A^{1}$ and $A^{2}$.

A Hamiltonian structure of these integrable hydrodynamic chains is unknown
at this moment. Just three Hamiltonian subcases have been found and
completely investigated. The Hamiltonian\ hydrodynamic chains (here $%
h_{2,k}\equiv \partial h_{2}/\partial A^{k},k=0,1,2$)%
\begin{equation*}
A_{t}^{k}=2h_{2,2}A_{x}^{k+1}+h_{2,1}A_{x}^{k}+(k+2)A^{k+1}(h_{2,2})_{x}+(k+1)A^{k}(h_{2,1})_{x}+kA^{k-1}(h_{2,0})_{x}
\end{equation*}%
are associated with the Kupershmidt--Manin Poisson bracket (see the second
part in \cite{FerMarsh}, \cite{GR} and \cite{KM})%
\begin{equation*}
\{A^{k}(x),A^{n}(x^{\prime })\}=[kA^{k+n-1}D_{x}+nD_{x}A^{k+n-1}]\delta
(x-x^{\prime }),\text{ \ }k,n=0,1,2,...;
\end{equation*}%
the Hamiltonian\ hydrodynamic chains (here $h_{1,k}\equiv \partial
h_{1}/\partial A^{k},k=0,1$)%
\begin{equation*}
A_{t}^{k}=(\alpha +\beta )h_{1,1}A_{x}^{k+1}+\beta h_{1,0}A_{x}^{k}+[\alpha
(k+1)+2\beta ]A^{k+1}(h_{1,1})_{x}+(\alpha k+2\beta )A^{k}(h_{1,0})_{x}
\end{equation*}%
are associated with the Kupershmidt Poisson brackets (see \cite{FKMP} and 
\cite{KuperNorm})%
\begin{equation*}
\{A^{k}(x),A^{n}(x^{\prime })\}=[(\alpha k+\beta )A^{k+n}D_{x}+(\alpha
n+\beta )D_{x}A^{k+n}]\delta (x-x^{\prime }),\text{ \ }k,n=0,1,2,...;
\end{equation*}%
the Hamiltonian\ hydrodynamic chains (see \cite{KuperPavl})%
\begin{equation*}
A_{t}^{k}=\beta h_{0}^{\prime }(A^{0})A_{x}^{k+1}+(\alpha k+2\beta
)A^{k+1}h_{0}^{\prime \prime }(A^{0})A_{x}^{0}
\end{equation*}%
are associated with the simplest Dorfman Poisson bracket (see \cite{Dorfman})%
\begin{equation*}
\{A^{k}(x),A^{n}(x^{\prime })\}=[(\alpha k+\beta )A^{k+n+1}D_{x}+(\alpha
n+\beta )D_{x}A^{k+n+1}]\delta (x-x^{\prime }),\text{ \ }k,n=0,1,2,...
\end{equation*}

If these hydrodynamic chains are integrable, then the corresponding
Hamiltonian densities $h_{0}(A^{0}),$ $h_{1}(A^{0},A^{1})$ and $%
h_{2}(A^{0},A^{1},A^{2})$ cannot be arbitrary. The simplest case $%
h_{0}(A^{0})$ is described in \cite{KuperPavl} (see also \cite{MaksHam}),
while a full list of admissible expressions $h_{1}(A^{0},A^{1})$ and $%
h_{2}(A^{0},A^{1},A^{2})$ is given in \cite{FKMP} and \cite{FerMarsh},
respectively.

Mainly, this paper is devoted to a description of integrable Hamiltonian
hydrodynamic chains%
\begin{equation}
A_{t}^{0}=2(\beta A^{1}+\delta A^{0}+\xi )h_{0}^{\prime \prime
}(A^{0})A_{x}^{0}+h_{0}^{\prime }(A^{0})(\beta A_{x}^{1}+\delta A_{x}^{0}),
\label{b}
\end{equation}%
\begin{equation*}
A_{t}^{k}=h_{0}^{\prime }(A^{0})(\beta A_{x}^{k+1}+\delta
A_{x}^{k})+[(\alpha k+2\beta )A^{k+1}+(\gamma k+2\delta )A^{k}+\epsilon
kA^{k-1}]h_{0}^{\prime \prime }(A^{0})A_{x}^{0},\text{ }k=1,2,...,
\end{equation*}%
associated with more general Dorfman Poisson brackets (see \cite{Dorfman})%
\begin{equation}
\{A^{k}(x),A^{n}(x^{\prime })\}=[\Gamma ^{kn}(\mathbf{A})D_{x}+D_{x}\Gamma
^{nk}(\mathbf{A})]\delta (x-x^{\prime }),\text{ \ \ \ }k,n=0,1,2,...
\label{poi}
\end{equation}%
where ($\alpha ,\beta ,\gamma ,\delta ,\epsilon ,\xi $ are arbitrary
constants)%
\begin{equation*}
\Gamma ^{00}=\beta A^{1}+\delta A^{0}+\xi ,\text{ \ }\Gamma ^{kn}=(\alpha
k+\beta )A^{k+n+1}+(\gamma k+\delta )A^{k+n}+\epsilon kA^{k+n-1},\text{ }%
k+n>0.
\end{equation*}%
Hamiltonian densities $h_{0}(A^{0})$ determining integrable hydrodynamic
chains (\ref{b}) are extracted by virtue of the moment decomposition
approach (see \cite{MaksHam}). The crucial feature of this moment
decomposition approach is a reducibility (i.e. the so called infinite
dimensional analogue of the Darboux theorem) of Dorfman Poisson brackets to
the canonical form $d/dx$.

This paper is organized in the following way. In Section \textbf{2}, the
first (and most general, i.e. all constants $\alpha ,\beta ,\gamma ,\delta
,\epsilon ,\xi $ are arbitrary) class of integrable hydrodynamic chains
associated with Dorfman Poisson brackets is described. Corresponding Vlasov
type kinetic equations are derived. Generating functions of conservation
laws and commuting flows are found. A hierarchy of such integrable
hydrodynamic chains is extended on negative moments and negative time
variables. Three dimensional two component hydrodynamic type systems (as
well as three dimensional quasilinear equations of the second order)
connected with this hierarchy are presented. In Section \textbf{3}, the
second class (i.e. $\beta =0$ but $\delta \neq 0$) of integrable
hydrodynamic chains is described. In comparison with the general case, these
hydrodynamic chains possess a momentum density, but Hamiltonian densities
depend on infinitely many moments $A^{k}$. In Section \textbf{4}, the third
(and the last, i.e. $\beta =0$ and $\delta =0$) class of integrable
hydrodynamic chains is described. In Conclusion, a relationship between
integrable Hamiltonian hydrodynamic chains and integrable Hamiltonian three
dimensional two component hydrodynamic type systems is discussed.

\section{Dorfman Poisson brackets and the moment decomposition approach}

The above infinite component Dorfman Poisson brackets can be reduced to the
canonical form (see \cite{dn})%
\begin{equation}
\{a^{i}(x),a^{k}(x^{\prime })\}=\frac{\delta _{ik}}{\epsilon _{i}}\delta
^{\prime }(x-x^{\prime }),\text{ \ \ }i,k=1,2,...,N,  \label{pou}
\end{equation}%
where $\epsilon _{i}$ are arbitrary constants ($\delta _{ik}$ is a Kronecker
symbol), due to the so-called \textquotedblleft moment
decomposition\textquotedblright 
\begin{equation}
A^{k}=\sum f_{k,i}(a^{i})  \label{dec}
\end{equation}%
with an appropriate choice of functions $f_{k,i}(a^{i})$ for \textit{any}
natural integer $N$.

\textbf{Theorem}: \textit{Suppose }$\beta \neq 0$,\textit{\ in such a
general case, Dorfman Poisson bracket} (\ref{poi}) \textit{reduces to }(\ref%
{pou}) \textit{under the moment decomposition}%
\begin{equation}
dA^{k}=\beta ^{-k}\sum \epsilon _{m}V(a^{m})(V^{\prime }(a^{m})-\delta
)^{k}da^{m},\text{ \ \ \ }k=0,1,...,  \label{mom}
\end{equation}%
\textit{where the function} $V(p)$ \textit{satisfies the ODE}%
\begin{equation}
VV^{\prime \prime }=\frac{\alpha }{\beta }V^{\prime ^{2}}+\left( \gamma -%
\frac{2\alpha \delta }{\beta }\right) V^{\prime }+\beta \epsilon -\gamma
\delta +\frac{\alpha \delta ^{2}}{\beta },  \label{sec}
\end{equation}%
\textit{and }$\xi $\textit{\ is an integration constant of the simplest
constraint}%
\begin{equation}
\frac{1}{2}\sum \epsilon _{m}V^{2}(a^{m})=\beta A^{1}+\delta A^{0}+\xi .
\label{strain}
\end{equation}

\textbf{Proof}: A substitution of moment decomposition (\ref{dec}) into
Poisson bracket (\ref{poi}) implies the recursive relationships%
\begin{equation*}
(\alpha k+\beta )f_{k+n+1,i}^{\prime }+(\gamma k+\delta )f_{k+n,i}^{\prime
}+\epsilon kf_{k+n-1,i}^{\prime }=\frac{1}{\epsilon _{i}}f_{n,i}^{\prime
}f_{k,i}^{\prime \prime },\text{ \ \ }k,n=0,1,2,...
\end{equation*}%
and constraint (\ref{strain}). For $k=0$, these ODE's%
\begin{equation*}
\beta f_{n+1,i}^{\prime }+\delta f_{n,i}^{\prime }=\frac{1}{\epsilon _{i}}%
f_{n,i}^{\prime }f_{0,i}^{\prime \prime },\text{ \ \ }n=0,1,2,...
\end{equation*}%
can be reduced to the common form ($\beta \neq 0$)%
\begin{equation*}
f_{n,i}^{\prime }=\frac{f_{0,i}^{\prime }}{\beta ^{n}}\left( \frac{%
f_{0,i}^{\prime \prime }}{\epsilon _{i}}-\delta \right) ^{n},\text{ \ \ }%
n=0,1,2,...
\end{equation*}%
A substitution of these expressions into the above recursive relationships
leads to (\ref{sec}), where $V(a^{i})=f_{0,i}^{\prime }/\epsilon ^{i}$.
Moreover, taking into account (\ref{sec}), an integration of the
differential $d(\alpha A^{k+2}+\gamma A^{k+1}+\epsilon A^{k})$ leads (see (%
\ref{mom})) to an infinite series of constraints%
\begin{equation*}
(\alpha k+2\beta )A^{k+1}+(\gamma k+2\delta )A^{k}+\epsilon kA^{k-1}+\xi
_{k}=\beta ^{-k}\sum \epsilon _{m}V^{2}(a^{m})(V^{\prime }(a^{m})-\delta
)^{k},\text{ \ }k=0,1,...,
\end{equation*}%
where $\xi _{k}$ are integration constants. It is easy to see, if $k=0$,
then $\xi _{0}=2\xi $ (see (\ref{strain})). The Theorem is proved.

For any positive integer $M$, an arbitrary Hamiltonian density $%
h_{M}(A^{0},A^{1},...,A^{M})$ and Poisson bracket (\ref{pou}) determine the
Hamiltonian hydrodynamic type system (see \cite{Tsar})%
\begin{equation}
a_{t}^{i}=\frac{1}{\epsilon _{i}}\partial _{x}\frac{\partial h_{M}}{\partial
a^{i}},  \label{sisa}
\end{equation}%
reducible to the symmetric form (see \cite{algebra}; here $h_{M,m}\equiv
\partial h/\partial A^{m},m=0,1,...,M$)%
\begin{equation}
a_{t}^{i}=\frac{1}{\epsilon _{i}}\partial _{x}\left( \overset{M}{\underset{%
m=0}{\sum }}h_{M,m}\frac{\partial A^{m}}{\partial a^{i}}\right) \equiv
\partial _{x}\left( V(a^{i})\overset{M}{\underset{m=0}{\sum }}\frac{%
(V^{\prime }(a^{i})-\delta )^{m}}{\beta ^{m}}h_{M,m}\right) .  \label{a}
\end{equation}%
It is easy to verify that (\ref{a}) is a hydrodynamic reduction of the
Hamiltonian hydrodynamic chain (associated with Dorfman Poisson bracket (\ref%
{poi}))%
\begin{eqnarray*}
A_{t}^{0} &=&2\xi (h_{M,0})_{x}+\overset{M}{\underset{n=0}{\sum }}[(\alpha
n+2\beta )A^{n+1}+(\gamma n+2\delta )A^{n}+\epsilon nA^{n-1}](h_{M,n})_{x} \\
&& \\
&&+\overset{M}{\underset{n=0}{\sum }}[(\alpha n+\beta )A_{x}^{n+1}+(\gamma
n+\delta )A_{x}^{n}+\epsilon nA_{x}^{n-1}]h_{M,n}, \\
&& \\
A_{t}^{k} &=&\overset{M}{\underset{n=0}{\sum }}[(\alpha (k+n)+2\beta
)A^{k+n+1}+(\gamma (k+n)+2\delta )A^{k+n}+\epsilon
(k+n)A^{k+n-1}](h_{M,n})_{x} \\
&& \\
&&+\overset{M}{\underset{n=0}{\sum }}[(\alpha n+\beta )A_{x}^{k+n+1}+(\gamma
n+\delta )A_{x}^{k+n}+\epsilon nA_{x}^{k+n-1}]h_{M,n}
\end{eqnarray*}%
by virtue of moment decomposition (\ref{mom}). If the Hamiltonian density $%
h_{M}(A^{0},A^{1},...,A^{M})$ is an arbitrary function, the above
hydrodynamic chain is non-integrable, as well as its hydrodynamic reduction (%
\ref{a}). Just in some special cases, this hydrodynamic chain and
hydrodynamic reduction (\ref{a}) became integrable. It means that the
Hamiltonian density $h_{M}(A^{0},A^{1},...,A^{M})$ must satisfy some
overdetermined system in partial derivatives, which can be obtained
utilizing different criteria of an integrability. For instance, in such a
case, Haantjes tensor vanishes (see detail in \cite{FerMarsh}), a family of
hydrodynamic reductions (see detail in \cite{FerKar}) parameterizes by $N$
arbitrary functions of a single variable. In this paper, integrable
hydrodynamic chains are extracted by the method of symmetric hydrodynamic
reductions (see detail in \cite{algebra}, \cite{MaksGen}, \cite{MaksHam}).
Without loss of generality, just the simplest Hamiltonian density $%
h_{0}(A^{0})$ is considered below. Indeed, if the Hamiltonian density $%
h_{0}(A^{0})$ determines integrable hydrodynamic chain (\ref{b}), then (\ref%
{b}) possesses an infinite series of conservation law densities $%
h_{n}(A^{0},A^{1},...,A^{n})$; then an infinite series of commuting flows
(determined by these Hamiltonian densities $h_{n}(A^{0},A^{1},...,A^{n})$)%
\begin{eqnarray}
A_{t^{n+1}}^{0} &=&2\xi (h_{n,0})_{x}+\overset{n}{\underset{m=0}{\sum }}%
[(\alpha m+2\beta )A^{m+1}+(\gamma m+2\delta )A^{m}+\epsilon
mA^{m-1}](h_{n,m})_{x}  \notag \\
&&  \notag \\
&&+\overset{n}{\underset{m=0}{\sum }}[(\alpha m+\beta )A_{x}^{m+1}+(\gamma
m+\delta )A_{x}^{m}+\epsilon mA_{x}^{m-1}]h_{n,m},  \notag \\
&&  \label{i} \\
A_{t^{n+1}}^{k} &=&\overset{n}{\underset{m=0}{\sum }}[(\alpha (k+m)+2\beta
)A^{k+m+1}+(\gamma (k+m)+2\delta )A^{k+m}+\epsilon
(k+m)A^{k+m-1}](h_{n,m})_{x}  \notag \\
&&  \notag \\
&&+\overset{n}{\underset{m=0}{\sum }}[(\alpha m+\beta )A_{x}^{k+m+1}+(\gamma
m+\delta )A_{x}^{k+m}+\epsilon mA_{x}^{k+m-1}]h_{n,m}  \notag
\end{eqnarray}%
exists. Since $M=0$, corresponding hydrodynamic reductions (\ref{a})
simplify to%
\begin{equation}
a_{t^{1}}^{i}=\partial _{x}(V(a^{i})h_{0}^{\prime }).  \label{reda}
\end{equation}

The integrability criterion contains three steps only (see detail in \cite%
{algebra}, \cite{MaksGen}, \cite{MaksHam} and \cite{OPS}).

\textbf{1}. Instead (\ref{reda}), let us introduce the generating function
of conservation laws for hydrodynamic type system (\ref{reda}) replacing $%
a^{i}$ by $p(x,t;\lambda )$, i.e.%
\begin{equation}
p_{t^{1}}=\partial _{x}(V(p)h_{0}^{\prime }),  \label{p}
\end{equation}%
where $\lambda $ is a free parameter (thus, $a^{i}=p(x,t,\lambda _{i})$,
where $\lambda _{i}$ are $N$ distinct values of the parameter $\lambda $).

\textbf{2}. Under the semi-hodograph transformation $p(x,t;\lambda
)\rightarrow \lambda (x,t;p)$, (\ref{p}) reduces to the Vlasov type kinetic
(the collisionless Boltzmann type) equation (see \cite{OPS})%
\begin{equation}
\lambda _{t^{1}}=V^{\prime }(p)h_{0}^{\prime }\lambda _{x}-V(p)\lambda
_{p}h_{0}^{\prime \prime }A_{x}^{0}.  \label{lya}
\end{equation}

\textbf{3}. Finally, one can check (see \cite{algebra}) the consistency (\ref%
{lya}) and corresponding hydrodynamic reduction (\ref{reda}).

Since, all moments $A^{k}$ are expressed via field variables $a^{n}$ (see (%
\ref{mom})), the last step can be replaced by a verification of consistency (%
\ref{b}) and (\ref{lya}). In such a case, all necessary computations (see 
\cite{FKMP} and \cite{FerMarsh}) significantly simplify. Indeed, a
substitution of $\lambda (p,\mathbf{A})$ into (\ref{lya}) leads to the
linear PDE system (here $\lambda _{p}\equiv \partial \lambda /\partial
p,\lambda _{k}\equiv \partial \lambda /\partial A^{k},k=0,1,...$)%
\begin{equation}
V\lambda _{p}+\overset{\infty }{\underset{k=0}{\sum }}[(\alpha k+2\beta
)A^{k+1}+(\gamma k+2\delta )A^{k}+\epsilon kA^{k-1}]\lambda _{k}+\left( 2\xi
+(\delta -V^{\prime })\frac{h_{0}^{\prime }}{h_{0}^{\prime \prime }}\right)
\lambda _{0}=0,  \label{c}
\end{equation}%
\begin{equation}
\lambda _{k}=\frac{\lambda _{0}}{q^{k}},\text{\ \ }k=0,1,2,...,  \label{d}
\end{equation}%
where%
\begin{equation}
q=\frac{V^{\prime }(p)-\delta }{\beta }.  \label{q}
\end{equation}%
A general solution of (\ref{d}) is given by%
\begin{equation}
\lambda (q,\mathbf{A})=B(q)\overset{\infty }{\underset{k=0}{\sum }}\frac{%
A^{k}}{q^{k}}+C(q),  \label{rim}
\end{equation}%
where $B(q)$ and $C(q)$ are not yet determined functions (\textquotedblleft
integration constants\textquotedblright ). A substitution (\ref{rim}) into (%
\ref{c}) leads to an extraction of an \textit{integrable} hydrodynamic chain
determined by the Hamiltonian density $h_{0}(A^{0})$ such that ($\sigma $ is
an arbitrary constant; $\alpha \neq 2\beta $)%
\begin{equation}
h_{0}^{\prime }=\left( A^{0}+\frac{\sigma }{\alpha -2\beta }\right) ^{\frac{%
\beta }{\alpha -2\beta }},  \label{hm}
\end{equation}%
while the coefficients $B(q)$ and $C(q)$ in an asymptotic expansion ($%
q\rightarrow \infty $) are given by their derivatives%
\begin{equation*}
(\ln B)^{\prime }=\frac{(\alpha -2\beta )q^{2}-2\delta q-\epsilon }{q(\alpha
q^{2}+\gamma q+\epsilon )},\text{\ \ }C^{\prime }=\frac{(\sigma q-2\xi )B}{%
\alpha q^{2}+\gamma q+\epsilon }.
\end{equation*}%
In a particular case $\alpha =2\beta $, an integrable hydrodynamic chain is
determined by the Hamiltonian density $h_{0}(A^{0})$ such that ($\sigma $ is
an arbitrary constant)%
\begin{equation*}
h_{0}^{\prime }=\exp (A^{0}/\sigma ),
\end{equation*}%
while the coefficients $B(q)$ and $C(q)$ in an asymptotic expansion ($%
q\rightarrow \infty $) are given by their derivatives%
\begin{equation*}
(\ln B)^{\prime }=-\frac{2\delta q+\epsilon }{q(2\beta q^{2}+\gamma
q+\epsilon )},\text{\ \ }C^{\prime }=\frac{(\beta \sigma q-2\xi )B}{2\beta
q^{2}+\gamma q+\epsilon }.
\end{equation*}

Equation (\ref{sec}) can be integrated in the parametric form (see (\ref{q}))%
\begin{equation}
V=\exp \int \frac{(\beta q+\delta )dq}{\alpha q^{2}+\gamma q+\epsilon },%
\text{\ \ \ }p=\int \frac{Vdq}{\alpha q^{2}+\gamma q+\epsilon }.  \label{exp}
\end{equation}%
Then all conservation law densities $h_{k}$ can be found by a substitution
of the inverse function $q(\lambda ,\mathbf{A})$ (expanded in the B\"{u}%
rmann--Lagrange series, see, for instance, \cite{ls}) in (\ref{exp}) at the
vicinity $q\rightarrow \infty $.

\subsection{Negative conservation laws and commuting flows}

All \textit{positive} commuting flows (\ref{i}) possess infinite sets of
conservation laws (cf. (\ref{e}))%
\begin{equation*}
\partial _{t^{n}}h_{k}=\partial _{x}f_{n,k}(h_{0},h_{1},...,h_{k+n}),\text{
\ }k=0,1,2,...,\text{ }n=2,3,...
\end{equation*}%
In the general case, integrable hydrodynamic chain (\ref{b}) possesses local
conservation laws for positive integers only. It means that any \textit{%
negative} conservation law density $h_{-k}$ must be expressed via \textit{%
infinitely many} \textit{negative} moments $A^{-n}$. Nevertheless, if $%
\epsilon =0$, each negative conservation law density $h_{-k}$ can be
expressed via a \textit{finite} number of negative moments $A^{-n}$.
However, $\epsilon =0$ is not a particular case. Indeed, let us introduce
new parameters $c_{k}$ such that $\alpha =\beta c_{2},\gamma =c_{1}+2\delta
c_{2},\beta \epsilon =c_{0}+c_{1}\delta +c_{2}\delta ^{2}$. Thus, instead
five arbitrary parameters $\alpha ,\beta ,\gamma ,\delta ,\epsilon $, we
shall use another five parameters $\beta ,\delta ,c_{0},c_{1},c_{2}$. In
such a case, (\ref{sec}) simplifies to (see \cite{MaksGen})%
\begin{equation}
VV^{\prime \prime }=c_{2}V^{\prime ^{2}}+c_{1}V^{\prime }+c_{0}.  \label{eq}
\end{equation}%
It means that integrable hydrodynamic chain (\ref{b}) is parameterized by
these three essential parameters only, because the parameter $\delta $ does
not exist in generating function of conservation laws (\ref{p}) as well as
in the Hamiltonian density (see (\ref{hm}))%
\begin{equation}
h_{0}^{\prime }=\left( A^{0}+\frac{\sigma }{\beta (c_{2}-2)}\right) ^{\frac{1%
}{c_{2}-2}},  \label{hom}
\end{equation}%
while the parameter $\beta $ can be incorporated to the integration constant 
$\sigma $. It is easy to see below, that the parameter $\beta $ can be
removed from Poisson bracket (\ref{poi})%
\begin{equation*}
\Gamma ^{kn}=\beta (c_{2}k+1)A^{k+n+1}+(k(c_{1}+2\delta c_{2})+\delta
)A^{k+n}+\frac{c_{0}+c_{1}\delta +c_{2}\delta ^{2}}{\beta }kA^{k+n-1}
\end{equation*}%
by an appropriate scaling of moments $A^{n}$. Thus, aforementioned
integrable hydrodynamic chains are determined by one-parametric family of
Dorfman Poisson brackets. Since $\delta $ is a free parameter, let us choose
the special value $\bar{\delta}$ of this parameter such that (i.e. $\epsilon
=0$)%
\begin{equation}
c_{0}+c_{1}\bar{\delta}+c_{2}\bar{\delta}^{2}=0.  \label{delta}
\end{equation}%
Then integrable hydrodynamic chain (\ref{b})

\begin{eqnarray*}
A_{t^{1}}^{0} &=&2(\beta A^{1}+\bar{\delta}A^{0}+\xi )(h_{0}^{\prime
})_{x}+h_{0}^{\prime }(\beta A_{x}^{1}+\bar{\delta}A_{x}^{0}), \\
&& \\
A_{t^{1}}^{k} &=&[\beta (c_{2}k+2)A^{k+1}+(k(c_{1}+2\bar{\delta}c_{2})+2\bar{%
\delta})A^{k}](h_{0}^{\prime })_{x}+h_{0}^{\prime }(\beta A_{x}^{k+1}+\bar{%
\delta}A_{x}^{k})
\end{eqnarray*}%
possesses an infinite series of negative local conservation laws

\begin{equation*}
\partial _{t^{1}}h_{-k}=\partial _{x}f_{k}(h_{0},h_{-1},...,h_{-k}),\text{ \ 
}k=0,1,2,...
\end{equation*}%
The first negative conservation law is given by%
\begin{equation}
\partial _{t^{1}}h_{-1}(A^{-1})=\bar{\delta}\partial
_{x}[h_{-1}(A^{-1})h_{0}^{\prime }(A^{0})],  \label{nach}
\end{equation}%
where%
\begin{equation*}
h_{-1}=\left( A^{-1}+\frac{\sigma }{2\bar{\delta}(1-c_{2})-c_{1}}\right) ^{%
\frac{\bar{\delta}}{2\bar{\delta}(1-c_{2})-c_{1}}}
\end{equation*}%
Thus, the first \textit{negative} commuting flow%
\begin{eqnarray*}
A_{t^{-1}}^{k} &=&h_{-1}^{\prime }[\beta (1-c_{2})A_{x}^{k}+(\bar{\delta}%
(1-2c_{2})-c_{1})A_{x}^{k-1}] \\
&& \\
&&+[\beta (c_{2}(k-1)+2)A^{k}+(k(c_{1}+2\bar{\delta}c_{2})+2\bar{\delta}%
(1-c_{2})-c_{1})A^{k-1}]\left( h_{-1}^{\prime }\right) _{x}
\end{eqnarray*}%
is determined by the above Hamiltonian density $h_{-1}(A^{-1})$. Then the
\textquotedblleft zeroth\textquotedblright\ conservation law is%
\begin{equation}
\partial _{t^{-1}}h_{0}(A^{0})=\beta (1-c_{2})\partial _{x}(h_{-1}^{\prime
}(A^{-1})h_{0}(A^{0})).  \label{alo}
\end{equation}

Under moment decomposition (\ref{mom}) extended to negative values of index $%
k$, the above hydrodynamic chain transforms to the hydrodynamic reduction%
\begin{equation*}
a_{t^{-1}}^{i}=\beta \partial _{x}\left( \frac{V(a^{i})}{V^{\prime }(a^{i})-%
\bar{\delta}}h_{-1}^{\prime }(A^{-1})\right)
\end{equation*}%
commuting with hydrodynamic type system (\ref{reda}). A corresponding
generating function of conservation laws is given by (cf. (\ref{p}))%
\begin{equation}
p_{t^{-1}}=\beta \partial _{x}\left( \frac{V(p)}{V^{\prime }(p)-\bar{\delta}}%
h_{-1}^{\prime }(A^{-1})\right) .  \label{s}
\end{equation}%
Thus, obviously, generating function of conservation laws (\ref{p}) is valid
for Hamiltonian densities depended on negative moments too. However, a
generating function of \textit{negative} conservation law densities is
associated with \textit{another} expansion (cf. (\ref{rim})) at the vicinity 
$q\rightarrow 0$. Indeed, a similar computation as in the previous
subsection leads to%
\begin{equation}
\lambda (q,\mathbf{A})=\tilde{B}(q)\overset{-1}{\underset{k=-\infty }{\sum }}%
q^{-k-1}A^{k}+\tilde{C}(q),  \label{t}
\end{equation}%
where%
\begin{equation*}
(\ln \tilde{B})^{\prime }=\frac{2\beta (c_{2}-1)q+c_{1}+2\bar{\delta}%
(c_{2}-1)}{\beta c_{2}q^{2}+(c_{1}+2\bar{\delta}c_{2})q},\text{\ \ }\tilde{C}%
^{\prime }=-\frac{\sigma \tilde{B}}{\beta c_{2}q^{2}+(c_{1}+2\bar{\delta}%
c_{2})q}.
\end{equation*}%
Thus, a generating function of conservation laws for the integrable
hydrodynamic chain determined by the Hamiltonian density $%
h_{M_{1},M_{2}}(A^{-M_{2}},...,A^{-1},A^{0},...,A^{M_{1}})$ is given by (cf.
(\ref{a}), (\ref{p}), (\ref{s}))%
\begin{equation*}
p_{t}=\partial _{x}\left( V(p)\overset{M_{1}}{\underset{m=-M_{2}}{\sum }}%
\frac{(V^{\prime }(p)-\bar{\delta})^{m}}{\beta ^{m}}h_{M_{1},M_{2},m}\right)
,
\end{equation*}%
where $h_{M_{1},M_{2},m}\equiv \partial h_{M_{1},M_{2}}/\partial
A^{m},m=0,\pm 1,\pm 2,...$

\textbf{Remark}: The hierarchy of integrable hydrodynamic chains determined
by the Hamiltonian densities $h_{-n}(A^{-n},...,A^{-1})$ and $%
h_{k}(A^{0},...,A^{k})$ transforms to itself due to $A^{n-1}\leftrightarrow
A^{-n}$ and $t^{n}\leftrightarrow t^{-n}$. Indeed, let us introduce an
auxiliary function $\tilde{V}(p)$ such that (see (\ref{s}))%
\begin{equation}
\tilde{V}(p)=\frac{V(p)}{V^{\prime }(p)-\bar{\delta}}.  \label{m}
\end{equation}%
This function $\tilde{V}(p)$ satisfies (cf. (\ref{eq}))%
\begin{equation*}
\tilde{V}\tilde{V}^{\prime \prime }=\tilde{c}_{2}\tilde{V}^{\prime ^{2}}+%
\tilde{c}_{1}\tilde{V}^{\prime }+\tilde{c}_{0},
\end{equation*}%
where%
\begin{equation*}
\tilde{c}_{2}=\frac{2\bar{\delta}c_{2}+c_{1}}{\bar{\delta}(2c_{2}-1)+c_{1}},%
\text{ \ }\tilde{c}_{1}=2(c_{2}-1)\frac{2\bar{\delta}c_{2}+c_{1}}{\bar{\delta%
}(2c_{2}-1)+c_{1}}-c_{2}
\end{equation*}%
\begin{equation*}
\tilde{c}_{0}=(c_{2}-1)^{2}\frac{2\bar{\delta}c_{2}+c_{1}}{\bar{\delta}%
(2c_{2}-1)+c_{1}}+c_{2}(1-c_{2}).
\end{equation*}%
Then an inverse transformation is given by%
\begin{equation}
V(p)=[\bar{\delta}(1-2c_{2})-c_{1}]\frac{\tilde{V}(p)}{\tilde{V}^{\prime
}(p)+c_{2}-1}.  \label{n}
\end{equation}%
It means, generating function of conservation laws (\ref{p}) under the
transformation $t^{1}\leftrightarrow t^{-1},A^{0}\leftrightarrow A^{-1}$
reduces to generating function of conservation laws (\ref{s}). Thus, taking
into account (\ref{m}) and (\ref{n}), all higher generating functions of
conservation laws%
\begin{equation}
p_{t^{k+1}}=\partial _{x}\left( V(p)\overset{k}{\underset{m=0}{\sum }}\frac{%
(V^{\prime }(p)-\bar{\delta})^{m}}{\beta ^{m}}h_{k,m}\right) ,\text{ \ }%
k=0,1,2,...  \label{gena}
\end{equation}%
transform to corresponding lower generating functions of conservation laws%
\begin{equation}
p_{t^{-n}}=\partial _{x}\left( V(p)\overset{-1}{\underset{m=-n}{\sum }}\frac{%
\beta ^{m}}{(V^{\prime }(p)-\bar{\delta})^{m}}h_{-n,m}\right) ,\text{ \ }%
n=1,2,...  \label{mena}
\end{equation}%
and vice versa.

\subsection{Generating function of commuting flows and conservation laws}

These generating functions (\ref{gena}) and (\ref{mena}) can be incorporated
in the sole generating function of conservation laws and commuting flows%
\begin{equation}
\partial _{\tau (\zeta )}p(\lambda )=\partial _{x}G(p(\lambda ),p(\zeta )),
\label{glob}
\end{equation}%
where the generating function of conservation law densities $p(\lambda )$ is
determined above, while an auxiliary function $p(\zeta )$ is obtained from $%
p(\lambda )$ replacing $\lambda $ by $\zeta $. Thus, a substitution of an
expansion $p(\zeta )$ at the vicinity $q\rightarrow \infty $ or $%
q\rightarrow 0$ into r.h.s. leads to generating functions (\ref{gena}) and (%
\ref{mena}) with an appropriate expansion of the so called \textquotedblleft
vertex\textquotedblright\ operator $\partial _{\tau (\zeta )}$ with respect
to parameter $\zeta $.

\textbf{Theorem}: \textit{The function} $G(p(\lambda ),p(\zeta ))$ \textit{%
is defined by the quadrature}%
\begin{equation*}
dG(p(\lambda ),p(\zeta ))=\left( Q(p(\zeta ))-\frac{V(p(\zeta ))R(p(\zeta ))%
}{V^{\prime }(p(\zeta ))-V^{\prime }(p(\lambda ))}\right) dp(\lambda )+\frac{%
V(p(\lambda ))R(p(\zeta ))}{V^{\prime }(p(\zeta ))-V^{\prime }(p(\lambda ))}%
dp(\zeta ),
\end{equation*}%
\textit{where}%
\begin{equation}
R(p(\zeta ))=\frac{V^{\prime \prime }(p(\zeta ))}{V(p(\zeta ))},\text{ \ \ \ 
}Q^{\prime }(p(\zeta ))=(c_{2}-1)R(p(\zeta )).  \label{zhuk}
\end{equation}

\textbf{Proof}: The compatibility condition $\partial _{t^{1}}(\partial
_{\tau (\zeta )}p(\lambda ))=\partial _{\tau (\zeta )}(\partial
_{t^{1}}p(\lambda ))$ (see (\ref{p})) implies%
\begin{equation}
\partial _{\tau (\zeta )}\ln h_{0}^{\prime }=Q(p(\zeta ))\partial _{x}\ln
h_{0}^{\prime }+R(p(\zeta ))\partial _{x}p(\zeta ),  \label{sik}
\end{equation}%
\begin{equation*}
\frac{\partial G(p(\lambda ),p(\zeta ))}{\partial p(\lambda )}=Q(p(\zeta ))-%
\frac{V(p(\zeta ))R(p(\zeta ))}{V^{\prime }(p(\zeta ))-V^{\prime }(p(\lambda
))},\text{ \ }\frac{\partial G(p(\lambda ),p(\zeta ))}{\partial p(\zeta )}=%
\frac{V(p(\lambda ))R(p(\zeta ))}{V^{\prime }(p(\zeta ))-V^{\prime
}(p(\lambda ))}
\end{equation*}%
where functions $Q(p(\zeta ))$ and $R(p(\zeta ))$ are not yet determined.
The compatibility condition%
\begin{equation*}
\frac{\partial }{\partial p(\lambda )}\frac{\partial G(p(\lambda ),p(\zeta ))%
}{\partial p(\zeta )}=\frac{\partial }{\partial p(\zeta )}\frac{\partial
G(p(\lambda ),p(\zeta ))}{\partial p(\lambda )}
\end{equation*}%
leads to (\ref{zhuk}). Theorem is proved.

If $c_{2}=1$, then $Q(p(\zeta ))$ is a removable constant due to admissible
shift in the vertex operator $\partial _{\tau (\zeta )}+$const$\partial
_{x}\rightarrow \partial _{\tau (\zeta )}$. Then (\ref{sik}) reduces to the
generating function of the \textquotedblleft zeroth\textquotedblright\
conservation laws for all (positive and negative) commuting flows 
\begin{equation*}
\partial _{\tau (\zeta )}\ln h_{0}^{\prime }=R(p(\zeta ))\partial
_{x}p(\zeta ).
\end{equation*}%
It is easy to see $h_{0}=-\ln h_{0}^{\prime }$ in agreement with (\ref{hom})
for $c_{2}=1$.

If $c_{2}\neq 1$, then (\ref{sik}) reduces to (see the second equation in (%
\ref{zhuk}))%
\begin{equation}
\partial _{\tau (\zeta )}[(h_{0}^{\prime })^{c_{2}-1}]=\partial
_{x}[Q(p(\zeta ))(h_{0}^{\prime })^{c_{2}-1}].  \label{key}
\end{equation}%
It is easy to see $h_{0}$ satisfies (\ref{hom}) for $c_{2}\neq 1$.

\subsection{Integrable three dimensional quasilinear equations of the second
order}

The method of hydrodynamic reductions (see detail in \cite{FerKar}) allows
to extract integrable three dimensional quasilinear equations of the second
order (see \cite{BFT}). In this subsection we present a list of some new
such equations associated with the hierarchy of commuting hydrodynamic
chains described above.

Compatibility conditions $(p_{t^{k}})_{t^{n}}=(p_{t^{n}})_{t^{k}}$ lead to
integrable three dimensional hydrodynamic type systems. If $k=1$ and $n=-1$,
then the three dimensional hydrodynamic type system (see the general case in 
\cite{MaksZiemek})%
\begin{equation*}
u_{t^{-1}}=(1-c_{2})vu_{x}-uv_{x},\text{ \ }v_{t^{1}}=\bar{\delta}%
uv_{x}+(c_{1}+(2c_{2}-1)\bar{\delta})vu_{x}
\end{equation*}%
is determined by the so called dispersionless Lax pair (see (\ref{p}) and (%
\ref{s}))%
\begin{equation}
p_{t^{1}}=\partial _{x}\left( V(p)u\right) ,\text{ \ \ }p_{t^{-1}}=\partial
_{x}\left( \frac{V(p)}{V^{\prime }(p)-\bar{\delta}}v\right) ,  \label{bara}
\end{equation}%
where $u=h_{0}^{\prime }(A^{0})$ and $v=h_{-1}^{\prime }(A^{-1})$. If $%
c_{2}\neq 1$ and $\tilde{c}_{2}\neq 1$ this hydrodynamic type system
possesses two local conservation laws (\ref{nach}) and (\ref{alo})%
\begin{equation}
(u^{c_{2}-1})_{t^{-1}}=(1-c_{2})(vu^{c_{2}-1})_{x},\text{ \ }(v^{\tilde{c}%
_{2}-1})_{t^{1}}=\bar{\delta}(uv^{\tilde{c}_{2}-1})_{x}.  \label{tri}
\end{equation}%
Then the integrable quasilinear three dimensional equation of the second
order (see \cite{BFT})%
\begin{equation}
z_{t^{1}t^{-1}}-\frac{z_{t^{-1}}}{z_{x}}z_{xt^{1}}=(z_{x})^{\frac{1}{c_{2}-1}%
}\left( \bar{\delta}z_{xt^{-1}}+\frac{\bar{\delta}c_{2}+c_{1}}{c_{2}-1}\frac{%
z_{t^{-1}}}{z_{x}}z_{xx}\right)   \label{dva}
\end{equation}%
can be obtain introducing the potential function $z$ such that $%
z_{x}=u^{c_{2}-1}$ and $z_{t^{-1}}=(1-c_{2})vu^{c_{2}-1}$. A similar
equation can be derived from the second conservation law by another
potential function $\tilde{z}$ such that $\tilde{z}_{x}=v^{\tilde{c}_{2}-1}$
and $\tilde{z}_{t^{1}}=\bar{\delta}uv^{\tilde{c}_{2}-1}$. In the particular
case $c_{2}=1$, (\ref{dva}) replaces by 
\begin{equation*}
z_{t^{1}t^{-1}}=e^{z_{x}}[\bar{\delta}z_{xt^{-1}}+(c_{1}+(2c_{2}-1)\bar{%
\delta})z_{t^{-1}}z_{xx}],
\end{equation*}%
while the first conservation law in (\ref{tri}) replaces by $(\ln
u)_{t^{-1}}=-v_{x}$. A substitution (where the potential function $\psi $ is
introduced for (\ref{bara}))%
\begin{equation*}
u=\frac{\psi _{t^{1}}}{V(\psi _{x})},\text{ \ \ \ }v=\frac{V^{\prime }(\psi
_{x})-\bar{\delta}}{V(\psi _{x})}\psi _{t^{-1}}
\end{equation*}%
into (\ref{tri}) implies the integrable quasilinear three dimensional
equation of the second order%
\begin{equation*}
\psi _{t^{1}t^{-1}}+(c_{1}+c_{2}\bar{\delta})\frac{V^{\prime }(\psi _{x})-%
\bar{\delta}}{V^{2}(\psi _{x})}\psi _{t^{1}}\psi _{t^{-1}}\psi
_{xx}=(1-c_{2})\frac{V^{\prime }(\psi _{x})-\bar{\delta}}{V(\psi _{x})}\psi
_{t^{-1}}\psi _{xt^{1}}+\frac{\bar{\delta}\psi _{t^{1}}}{V(\psi _{x})}\psi
_{xt^{-1}}.
\end{equation*}

If $k=1$ and $n=2$, the dispersionless Lax pair (see (\ref{p}) and (\ref%
{gena}) where $M=1$)%
\begin{equation}
p_{t^{1}}=\partial _{x}\left( V(p)u\right) ,\text{ \ \ }p_{t^{2}}=\partial
_{x}\left( V(p)w+V(p)V^{\prime }(p)s\right)   \label{para}
\end{equation}%
determines the three dimensional hydrodynamic type system ($c_{2}\neq -1$)
written in the conservative form%
\begin{equation}
(u^{c_{2}-1})_{t^{1}}=\frac{c_{2}-1}{c_{2}+1}\left( \frac{w}{u}-\frac{c_{1}}{%
c_{2}}u^{c_{2}}\right) _{x},\text{ \ \ }u_{t^{2}}=w_{t^{1}}+\frac{c_{0}}{%
c_{2}+2}(u^{c_{2}+2})_{x},  \label{uk}
\end{equation}%
where $u=h_{0}^{\prime }(A^{0}),w=h_{1,0}-\bar{\delta}h_{1,1},s=h_{1,1}$.
Moreover, the compatibility condition $%
(p_{t^{1}})_{t^{2}}=(p_{t^{2}})_{t^{1}}$ leads to the constraint $%
s=u^{c_{2}+1}$, which precisely coincides with a relationship between the
corresponding Hamiltonian densities $h_{0}(A^{0})$ and $h_{1}(A^{0},A^{1})$,
i.e. $h_{1,1}=(h_{0}^{\prime })^{c_{2}+1}$. Then the quasilinear three
dimensional equation of the second order%
\begin{equation}
\frac{1}{c_{2}+1}z_{xt^{2}}=z_{x}z_{t^{1}t^{1}}+\left( \frac{z_{t^{1}}}{%
c_{2}-1}+\frac{c_{1}}{c_{2}}(z_{x})^{\frac{c_{2}}{c_{2}-1}}\right)
z_{xt^{1}}+\frac{c_{0}}{c_{2}+1}(z_{x})^{\frac{c_{2}+1}{c_{2}-1}}z_{xx},
\label{odin}
\end{equation}%
is associated with the same (see (\ref{dva})) potential function $z$ of the
first conservation law in (\ref{uk}). In the particular case $c_{2}=1$, (\ref%
{odin}) replaces by%
\begin{equation*}
z_{xt^{2}}=2z_{t^{1}t^{1}}+2(z_{t^{1}}+c_{1}e^{z_{x}})z_{xt^{1}}+c_{0}e^{2z_{x}}z_{xx},
\end{equation*}%
while the first conservation law in (\ref{uk}) replaces by 
\begin{equation*}
(\ln u)_{t^{1}}=\left( \frac{w}{2u}-\frac{c_{1}u}{2}\right) _{x}.
\end{equation*}%
A substitution (where the potential function $\psi $ is introduced for (\ref%
{para}); see also (\ref{bara}))%
\begin{equation*}
u=\frac{\psi _{t^{1}}}{V(\psi _{x})},\text{ \ \ \ }w=\frac{\psi _{t^{2}}}{%
V(\psi _{x})}-V^{\prime }(\psi _{x})\left( \frac{\psi _{t^{1}}}{V(\psi _{x})}%
\right) ^{c_{2}+1}
\end{equation*}%
into (\ref{uk}) implies the integrable quasilinear three dimensional
equation of the second order%
\begin{equation*}
\psi _{xt^{2}}=\left[ \frac{\psi _{t^{2}}}{\psi _{t^{1}}}+\left( \frac{\psi
_{t^{1}}}{V(\psi _{x})}\right) ^{c_{2}}(c_{1}-V^{\prime }(\psi _{x}))\right]
\psi _{xt^{1}}+(c_{2}+1)\left( \frac{\psi _{t^{1}}}{V(\psi _{x})}\right)
^{c_{2}-1}\psi _{t^{1}t^{1}}+c_{0}\left( \frac{\psi _{t^{1}}}{V(\psi _{x})}%
\right) ^{c_{2}+1}\psi _{xx}
\end{equation*}

The most interesting and exceptional case is given by $c_{2}=-1$. In such a
case, dispersionless Lax pair (\ref{para}) is no longer correct. The
dispersionless Lax pair%
\begin{equation}
p_{t^{1}}=\partial _{x}\left( V(p)u\right) ,\text{ \ \ }p_{t^{2}}=\partial
_{x}\left( V(p)w+V(p)V^{\prime }(p)s+V(p)F(V^{\prime }(p))\right) ,
\label{fu}
\end{equation}%
where $F(q)$ satisfies%
\begin{equation}
qF(q)+(-q^{2}+c_{1}q+c_{0})F^{\prime }(q)=q^{2},  \label{rya}
\end{equation}%
determines the pair of conservation laws (cf. (\ref{uk}))%
\begin{equation}
(u^{-2})_{t^{1}}+2\left( \frac{w+c_{1}\ln u+c_{1}}{u}\right) _{x}=0,\text{ \ 
}u_{t^{2}}=w_{t^{1}}+c_{0}[u(\ln u-1)]_{x}.  \label{f}
\end{equation}%
Then the quasilinear three dimensional equation of the second order%
\begin{equation*}
z_{xt^{2}}=z_{x}z_{t^{1}t^{1}}-\left( \frac{z_{t^{1}}}{2}+c_{1}(z_{x})^{1/2}%
\right) z_{xt^{1}}-\frac{c_{0}}{2}\ln z_{x}\cdot z_{xx}
\end{equation*}%
follows from the second conservation law of (\ref{f}), while $z$ is a
potential function of the first conservation law of (\ref{f}).

A comparison (\ref{fu}) with (\ref{gena}) means that the Hamiltonian density 
$h_{1}$ depends on $A^{0},A^{1}$ (as usual in the general case) and \textit{%
linearly} on \textit{all} other positive moments $A^{k}$ (see (\ref{i})). A
corresponding expression for the function $F(q)$ can be found by a
substitution of the Taylor series $F(q)=q^{3}(\epsilon _{3}+\epsilon
_{4}q+\epsilon _{5}q^{2}+...)$ into (\ref{rya}). Then all coefficients are
determined iteratively, i.e.%
\begin{equation*}
\epsilon _{3}=\frac{1}{3c_{0}},\text{ \ \ }\epsilon _{4}=-\frac{c_{1}}{%
4c_{0}^{2}},\text{ \ \ }\epsilon _{k}=\frac{(k-3)\epsilon
_{k-2}-(k-1)c_{1}\epsilon _{k-1}}{kc_{0}},\text{ \ }k=5,6,...
\end{equation*}

The next integrable three dimensional quasilinear equation of the second
order follows from (\ref{p}) and (\ref{key})%
\begin{equation*}
\partial _{\tau }(u^{c_{2}-1})=\partial _{x}(Q(p)u^{c_{2}-1}),\text{ \ \ }%
\partial _{t^{1}}p=\partial _{x}(V(p)u).
\end{equation*}%
Introducing a potential function $z$ such that (utilizing the first
conservation law) $z_{x}=u^{c_{2}-1},z_{\tau }=Q(p)z_{x}$, the second
conservation law transforms to%
\begin{equation*}
z_{x}z_{\tau t^{1}}-z_{\tau }z_{xt^{1}}=\frac{1}{c_{2}-1}\frac{V(p(s))}{%
p^{\prime }(s)}(z_{x})^{\frac{c_{2}}{c_{2}-1}}z_{xx}+(z_{x})^{\frac{1}{%
c_{2}-1}}V^{\prime }(p(s))(z_{x}z_{\tau x}-z_{\tau }z_{xx}),
\end{equation*}%
where $p(s)$ is an inverse function to $Q(p)$ and $s=z_{\tau }/z_{x}$.
Introducing a potential function $\psi $ such that (utilizing the second
conservation law) $\psi _{x}=p,\psi _{t^{1}}=V(p)u$, the first conservation
law transforms to another integrable three dimensional quasilinear equation
of the second order%
\begin{equation*}
\frac{\psi _{t^{1}\tau }}{\psi _{t^{1}}}-\frac{V^{\prime }(\psi _{x})}{%
V(\psi _{x})}\psi _{x\tau }=Q(\psi _{x})\frac{\psi _{xt^{1}}}{\psi _{t^{1}}}%
+\left( \frac{Q^{\prime }(\psi _{x})}{c_{2}-1}-Q(\psi _{x})\frac{V^{\prime
}(\psi _{x})}{V(\psi _{x})}\right) \psi _{xx}.
\end{equation*}

The most complicated integrable three dimensional quasilinear equation of
the second order associated with an integrable hierarchy of commuting
hydrodynamic chains presented in this Section (see (\ref{gena}) and (\ref%
{mena})) can be obtained utilizing (\ref{glob}). The compatibility condition
of two copies of (\ref{glob})%
\begin{equation*}
\partial _{\tau ^{1}}p(\lambda )=\partial _{x}G(p(\lambda ),p^{1}),\text{ \
\ \ }\partial _{\tau ^{2}}p(\lambda )=\partial _{x}G(p(\lambda ),p^{2})
\end{equation*}%
leads to the pair of conservation laws%
\begin{equation*}
\partial _{\tau ^{1}}p^{2}=\partial _{x}G(p^{2},p^{1}),\text{ \ \ \ }%
\partial _{\tau ^{2}}p^{1}=\partial _{x}G(p^{1},p^{2}),
\end{equation*}%
where $\partial _{\tau ^{1}}=\partial _{\tau (\zeta )},\partial _{\tau
^{2}}=\partial _{\tau (\eta )},p^{1}=p(\zeta ),p^{2}=p(\eta )$ and $\eta $
is an auxiliary parameter such $\zeta $. Let us introduce two potential
functions $\psi ^{1}$ and $\psi ^{2}$ such that $\psi _{x}^{1}=p^{1},\psi
_{\tau ^{1}}^{1}=G(p^{1},p^{2})$ and $\psi _{x}^{2}=p^{2},\psi _{\tau
^{1}}^{2}=G(p^{2},p^{1})$, then the above integrable three dimensional
hydrodynamic type system reduces to two equivalent three dimensional
quasilinear equations of the second order%
\begin{equation*}
\partial _{\tau ^{1}}\tilde{G}(\psi _{\tau ^{2}}^{1},\psi _{x}^{1})=\partial
_{x}G(\tilde{G}(\psi _{\tau ^{2}}^{1},\psi _{x}^{1}),\psi _{x}^{1}),\text{ \
\ \ }\partial _{\tau ^{2}}\tilde{G}(\psi _{\tau ^{1}}^{2},\psi
_{x}^{2})=\partial _{x}G(\tilde{G}(\psi _{\tau ^{1}}^{2},\psi _{x}^{2}),\psi
_{x}^{2}),
\end{equation*}%
where $p^{1}=\tilde{G}(\psi _{\tau ^{1}}^{2},\psi _{x}^{2})$ and $p^{2}=%
\tilde{G}(\psi _{\tau ^{2}}^{1},\psi _{x}^{1})$.

Under the transformation of independent variables $x\leftrightarrow t$,
generating function of conservation laws (\ref{p})%
\begin{equation*}
p_{x}=\partial _{t}(V(p)u)
\end{equation*}%
reduces to the form%
\begin{equation*}
q_{t}=\partial _{x}f\left( \frac{q}{u}\right) ,
\end{equation*}%
where $p=f(s),s=V(p)$ and $us=q$. In the particular case $c_{0}=0$, (\ref{eq}%
) reduces to the form (i.e. $c_{2}=c,c_{1}=1$)%
\begin{equation*}
V(p)V^{\prime \prime }(p)=cV^{\prime ^{2}}(p)+V^{\prime }(p)
\end{equation*}%
by an appropriate scaling of the independent variable $p$. Then the
dispersionless Lax pair%
\begin{equation}
q_{t}=\partial _{x}f\left( \frac{q}{u}\right) ,\text{ \ \ }q_{y}=\partial
_{x}f\left( \frac{q}{a}\right) ,  \label{conp}
\end{equation}%
determines the three dimensional hydrodynamic type system%
\begin{equation*}
a_{t}=\partial _{x}f\left( \frac{a}{u}\right) ,\text{ \ \ \ }u_{y}=\partial
_{x}f\left( \frac{u}{a}\right) ,
\end{equation*}%
where%
\begin{equation*}
f^{\prime }(s)=\frac{c}{s^{c}-1}.
\end{equation*}%
Also this three dimensional hydrodynamic type system%
\begin{equation}
a_{t}=\frac{u^{c-2}}{u^{c}-a^{c}}(au_{x}-ua_{x}),\text{ \ \ \ }u_{y}=\frac{%
a^{c-2}}{u^{c}-a^{c}}(au_{x}-ua_{x})  \label{corp}
\end{equation}%
can be rewritten as the three dimensional quasilinear equation of the second
order (see the general case in \cite{BFT})%
\begin{equation*}
Z_{xy}=c^{2}\frac{V^{2c}(Z_{t})}{(V^{c}(Z_{t})-1)^{2}}Z_{xt}+\frac{%
V^{c}(Z_{t})-1}{cV(Z_{t})}Z_{x}Z_{yt},
\end{equation*}%
where $a=Z_{x}$ and $u=Z_{x}/V(Z_{t})$. Introducing a potential function $%
\psi $ such that (see (\ref{conp})) $q=\psi _{x},u=\psi _{x}/V(\psi
_{t}),a=\psi _{x}/V(\psi _{y})$, (\ref{corp}) reduces to the semi-symmetric
form (i.e. this equation is invariant with respect to $y\leftrightarrow t$;
see the general case in \cite{BFT})%
\begin{equation*}
\psi _{yt}=c\frac{V(\psi _{y})V^{c}(\psi _{t})\psi _{xt}-V(\psi
_{t})V^{c}(\psi _{y})\psi _{xy}}{V^{c}(\psi _{t})-V^{c}(\psi _{y})}.
\end{equation*}%
This equation can be written in the form%
\begin{equation*}
\frac{V^{c}(\psi _{t})-V^{c}(\psi _{y})}{cV(\psi _{t})V(\psi _{y})}\psi
_{yt}=V^{c-1}(\psi _{t})\psi _{xt}-V^{c-1}(\psi _{y})\psi _{xy}.
\end{equation*}%
Let us take two extra copies of this equation%
\begin{eqnarray*}
\frac{V^{c}(\psi _{y})-V^{c}(\psi _{\tau })}{cV(\psi _{y})V(\psi _{\tau })}%
\psi _{y\tau } &=&V^{c-1}(\psi _{y})\psi _{xy}-V^{c-1}(\psi _{\tau })\psi
_{x\tau }, \\
&& \\
\frac{V^{c}(\psi _{\tau })-V^{c}(\psi _{t})}{cV(\psi _{\tau })V(\psi _{t})}%
\psi _{t\tau } &=&V^{c-1}(\psi _{\tau })\psi _{x\tau }-V^{c-1}(\psi
_{t})\psi _{xt},
\end{eqnarray*}%
where $\tau $ is the \textquotedblleft fourth time\textquotedblright\
variable. Eliminating second derivatives $\psi _{x\tau },\psi _{xt},\psi
_{xy}$, one can obtain the remarkable three dimensional equation of the
second order (see \cite{BFT})%
\begin{equation*}
\frac{V^{c}(\psi _{t})-V^{c}(\psi _{y})}{V(\psi _{t})V(\psi _{y})}\psi _{yt}+%
\frac{V^{c}(\psi _{y})-V^{c}(\psi _{\tau })}{V(\psi _{y})V(\psi _{\tau })}%
\psi _{y\tau }+\frac{V^{c}(\psi _{\tau })-V^{c}(\psi _{t})}{V(\psi _{\tau
})V(\psi _{t})}\psi _{t\tau }=0.
\end{equation*}

\section{The first degenerate level $\protect\beta =0$}

In Section \textbf{2}, the Theorem was formulated for the general case $%
\beta \neq 0$. This Section is devoted to this degeneration $\beta =0$.

\textbf{Theorem}: \textit{Suppose }$\beta =0$, \textit{but} $\delta \neq 0$,%
\textit{\ in such a case, Dorfman Poisson bracket} (\ref{poi}) \textit{%
reduces to }(\ref{pou}) \textit{under the moment decomposition}%
\begin{equation}
dA^{k}=\delta \sum \epsilon _{m}a^{m}(W^{\prime }(a^{m}))^{k}da^{m},\text{ \
\ \ }k=0,1,...,  \label{ak}
\end{equation}%
\textit{where the function} $W(p)$ \textit{satisfies the ODE}%
\begin{equation}
\delta pW^{\prime \prime }=\alpha W^{\prime ^{2}}+\gamma W^{\prime
}+\epsilon ,  \label{rel}
\end{equation}%
\textit{and }$\xi $\textit{\ is an integration constant of the simplest
constraint}%
\begin{equation}
\frac{\delta ^{2}}{2}\sum \epsilon _{m}(a^{m})^{2}=\delta A^{0}+\xi .
\label{atio}
\end{equation}

\textbf{Proof}: Under the substitution $V=\beta W+\delta p$, (\ref{sec})
reduces to the form%
\begin{equation*}
\beta WW^{\prime \prime }+\delta pW^{\prime \prime }=\alpha W^{\prime
^{2}}+\gamma W^{\prime }+\epsilon .
\end{equation*}%
Thus, the degenerate case $\beta =0$ is associated with (\ref{rel}). A
substitution of moment decomposition (\ref{dec}) into the Dorfman Poisson
bracket (cf. (\ref{poi}))%
\begin{equation*}
\{A^{k}(x),A^{n}(x^{\prime })\}=[\Gamma ^{kn}(\mathbf{A})D_{x}+D_{x}\Gamma
^{nk}(\mathbf{A})]\delta (x-x^{\prime }),\text{ \ \ \ }k,n=0,1,2,...
\end{equation*}%
where ($\alpha ,\gamma ,\delta ,\epsilon ,\xi $ are arbitrary constants)%
\begin{equation}
\Gamma ^{00}=\delta A^{0}+\xi ,\text{ \ }\Gamma ^{kn}=\alpha
kA^{k+n+1}+(\gamma k+\delta )A^{k+n}+\epsilon kA^{k+n-1},\text{ }k+n>0,
\label{poj}
\end{equation}%
implies the recursive relationships%
\begin{equation*}
\alpha nW_{k+n+1}+(\gamma n+\delta )W_{k+n}+\epsilon
nW_{k+n-1}=W_{k}W_{n}^{\prime },\text{ \ \ }k,n=0,1,2,...,
\end{equation*}%
where $W_{n}(a^{i})=f_{n,i}^{\prime }/\epsilon _{i}$. Following the general
case (see (\ref{mom}), (\ref{sec}) and below), suppose that a solution of
this system is given by $W_{m}=W_{0}(W^{\prime })^{m}$, where $W_{0}(p)$ and 
$W(p)$ are not yet determined. If $n=0$, then $W_{0}(p)=\delta p$ (up to an
additive constant). A substitution of the above ansatz into the recursive
relationships leads to (\ref{rel}). Moreover, taking into account (\ref{rel}%
), an integration of the differential $d(\alpha A^{k+2}+\gamma
A^{k+1}+\epsilon A^{k})$ leads (see (\ref{ak})) to an infinite series of
constraints%
\begin{equation*}
\alpha kA^{k+1}+(\gamma k+2\delta )A^{k}+\epsilon kA^{k-1}+\xi _{k}=\delta
^{2}\sum \epsilon _{m}(a^{m})^{2}(W^{\prime }(a^{m}))^{k},\text{ \ }%
k=0,1,...,
\end{equation*}%
where $\xi _{k}$ are integration constants. It is easy to see, if $k=0$,
then $\xi _{0}=2\xi $ (see (\ref{atio})). The Theorem is proved.

For any positive integer $M$, an arbitrary Hamiltonian density $%
h_{M}(A^{0},A^{1},...,A^{M})$ and Dorfman Poisson bracket (\ref{poj})
determine a hydrodynamic chain (see the previous Section), whose Hamiltonian
hydrodynamic reduction (\ref{sisa}) is presented in the symmetric form (see (%
\ref{ak}) and \cite{algebra}; here $h_{M,m}\equiv \partial h/\partial
A^{m},m=0,1,...,M$)%
\begin{equation*}
a_{t}^{i}=\frac{1}{\epsilon _{i}}\partial _{x}\left( \overset{M}{\underset{%
m=0}{\sum }}h_{M,m}\frac{\partial A^{m}}{\partial a^{i}}\right) \equiv
\delta \partial _{x}\left( a^{i}\overset{M}{\underset{m=0}{\sum }}(W^{\prime
}(a^{i}))^{m}h_{M,m}\right) .
\end{equation*}%
Then such an integrable hydrodynamic chain possesses the generating function
of conservation laws (cf. (\ref{gena}))%
\begin{equation}
p_{t}=\delta \partial _{x}\left( p\overset{M}{\underset{m=0}{\sum }}%
(W^{\prime }(p))^{m}h_{M,m}\right) .  \label{gen}
\end{equation}%
In this case, the Hamiltonian density satisfies some overdetermined system.
Following the general approach presented in the previous Section, we would
like restrict our consideration to the simplest case $h_{0}(A^{0})$ only.
However, this is impossible. In comparison with full ($\beta \neq 0$)
Dorfman Poisson bracket (\ref{poi}), this reduced ($\beta =0$) Dorfman
Poisson bracket leads to the momentum density $P=A^{0}$, i.e. an arbitrary
Hamiltonian hydrodynamic chain possesses a conservation law of the momentum 
\begin{equation*}
A_{t}^{0}=\left( 2\xi h_{M,0}+\alpha \overset{M}{\underset{n=0}{\sum }}%
nh_{M,n}A^{n+1}+\overset{M}{\underset{n=0}{\sum }}(\gamma n+2\delta
)h_{M,n}A^{n}+\epsilon \overset{M}{\underset{n=0}{\sum }}nh_{M,n}A^{n-1}-%
\delta h_{M}\right) _{x}.
\end{equation*}%
It looks like the simplest integrable case is determined by the next
Hamiltonian density $h_{1}(A^{0},A^{1})$. However, corresponding
hydrodynamic chain%
\begin{eqnarray*}
A_{t}^{0} &=&\left( 2\xi h_{1,0}+[\alpha A^{2}+(\gamma +2\delta
)A^{1}+\epsilon A^{0}]h_{1,1}+2\delta A^{0}h_{1,0}-\delta h_{1}\right) _{x},
\\
&& \\
A_{t}^{k} &=&[\alpha (k+1)A^{k+2}+(\gamma (k+1)+2\delta )A^{k+1}+\epsilon
(k+1)A^{k}](h_{1,1})_{x}
\end{eqnarray*}%
\begin{equation*}
+[\alpha kA^{k+1}+(\gamma k+2\delta )A^{k}+\epsilon
kA^{k-1}](h_{1,0})_{x}+h_{1,1}[\alpha A_{x}^{k+2}+(\gamma +\delta
)A_{x}^{k+1}+\epsilon A_{x}^{k}]+\delta h_{1,0}A_{x}^{k}
\end{equation*}%
depends on the highest moment $A^{k+2}$. Moreover, any higher commuting flow
determined by the Hamiltonian density $h_{n}(A^{0},...,A^{n})$ depends on
the highest moment $A^{k+n+1}$. Corresponding \textquotedblleft
time\textquotedblright\ variable $t^{n+1}$ changes from $n=1$, while $t^{0}$
must be reserved \textit{apriori }for $x$, because the \textquotedblleft
zeroth\textquotedblright\ conservation law density $h_{0}(A^{0})$ is a
momentum density $A^{0}$, which cannot create an \textquotedblleft
intermediate\textquotedblright\ hydrodynamic chain containing the highest
moment $A^{k+1}$. The momentum density generates just \textquotedblleft
trivial\textquotedblright\ commuting flow, i.e. $A_{t^{0}}^{k}=A_{x}^{k}$.
This is a reason to identify $x$ and $t^{0}$. Nevertheless, a commuting flow
determined by the \textquotedblleft time\textquotedblright\ variable $t^{1}$
exists.

\textbf{Lemma}: \textit{The hydrodynamic chain}%
\begin{equation}
A_{t^{1}}^{0}=[2(\xi +\delta A^{0})f^{\prime \prime }+\delta f^{\prime
}+\epsilon b_{1}]A_{x}^{0}+[(\gamma +\delta )b_{1}+2\epsilon b_{2}]A_{x}^{1},
\label{pyat}
\end{equation}%
\begin{equation*}
A_{t^{1}}^{k}=[\alpha kA^{k+1}+(\gamma k+2\delta )A^{k}+\epsilon
kA^{k-1}]f^{\prime \prime }A_{x}^{0}+(\delta f^{\prime }+\epsilon
b_{1})A_{x}^{k}+[(\gamma +\delta )b_{1}+2\epsilon b_{2}]A_{x}^{k+1},\text{ \ 
}k=1,2,...
\end{equation*}%
\textit{is determined by the Hamiltonian density depended nonlinearly on the
\textquotedblleft zeroth\textquotedblright\ moment and linearly on \textbf{%
all} higher moments, i.e.}%
\begin{equation}
h_{\ast }=f(A^{0})+\underset{n=1}{\overset{\infty }{\sum }}b_{n}A^{n},
\label{ham}
\end{equation}%
\textit{where} $b_{1}$ \textit{and} $b_{2}$ \textit{are arbitrary constants,
while all other constants} $b_{n}$ \textit{satisfy the recursive
relationships}%
\begin{equation}
\alpha (n-1)b_{n-1}+(\gamma n+\delta )b_{n}+\epsilon (n+1)b_{n+1}=0,\text{ \ 
}n=2,3,...  \label{rek}
\end{equation}

\textbf{Proof}: can obtained by a straightforward calculation.

Moreover, integrable hydrodynamic chain (\ref{pyat}) is associated with the
generating function of conservation laws (see (\ref{gen}))%
\begin{equation}
p_{t^{1}}=\delta \partial _{x}(pf^{\prime }(A^{0})+\bar{W}(p)),  \label{sest}
\end{equation}%
where%
\begin{equation}
\bar{W}(p)=p\overset{\infty }{\underset{m=1}{\sum }}b_{m}(W^{\prime
}(p))^{m}.  \label{z}
\end{equation}%
In this case (see \cite{OPS}),%
\begin{equation}
p\bar{W}^{\prime \prime }=c_{2}\bar{W}^{\prime ^{2}}+c_{1}\bar{W}^{\prime
}+c_{0},  \label{svya}
\end{equation}%
where $c_{k}$ are some constants.

\textbf{Theorem}: \textit{Integrable hydrodynamic chain} (\ref{pyat}) 
\textit{is determined by Hamiltonian density} (\ref{ham}), \textit{where}%
\begin{equation}
f^{\prime \prime }(A^{0})=\frac{(\gamma +\delta )b_{1}+2\epsilon b_{2}}{%
\alpha A^{0}+\sigma \lbrack (\gamma +\delta )b_{1}+2\epsilon b_{2}]},
\label{fun}
\end{equation}%
$\sigma $ \textit{is an integration constant, and constants} $b_{k}$ \textit{%
satisfy }(\ref{rek}).

\textbf{Proof}: Under the semi-hodograph transformation $p(\lambda
,x,t)\rightarrow \lambda (p,x,t)$, (\ref{sest}) reduces to the Vlasov type
kinetic equation (see \cite{OPS})%
\begin{equation*}
\lambda _{t^{1}}=\delta \lbrack (f^{\prime }(A^{0})+\bar{W}^{\prime
}(p))\lambda _{x}-p\lambda _{p}(f^{\prime })_{x}].
\end{equation*}%
A substitution $\lambda (p,\mathbf{A})$ leads to%
\begin{equation}
\delta p\lambda _{p}=\left[ \frac{(\gamma +\delta )b_{1}+2\epsilon b_{2}}{%
f^{\prime \prime }(A^{0})}q-2\xi -\overset{\infty }{\underset{k=0}{\sum }}%
\frac{\alpha kA^{k+1}+(\gamma k+2\delta )A^{k}+\epsilon kA^{k-1}}{q^{k}}%
\right] \lambda _{0}  \label{g}
\end{equation}%
and (\ref{d}), where (instead (\ref{q}))%
\begin{equation}
q=\frac{\delta \bar{W}^{\prime }(p)-\epsilon b_{1}}{(\gamma +\delta
)b_{1}+2\epsilon b_{2}}.  \label{quk}
\end{equation}%
Taking into account (see (\ref{svya}) and (\ref{quk}))%
\begin{equation}
\frac{\delta ^{2}p\bar{W}^{\prime \prime }(p)}{(\gamma +\delta
)b_{1}+2\epsilon b_{2}}=\tilde{c}_{2}q^{2}+\tilde{c}_{1}q+\tilde{c}_{0},
\label{second}
\end{equation}%
a substitution (\ref{rim}) into (\ref{g}) implies the constraints $\tilde{c}%
_{2}=\alpha ,\tilde{c}_{1}=\gamma ,\tilde{c}_{0}=\epsilon $, (\ref{fun}) and%
\begin{equation}
(\ln B)^{\prime }=\frac{\alpha q^{2}-2\delta q-\epsilon }{q(\alpha
q^{2}+\gamma q+\epsilon )},\text{ \ \ }C^{\prime }=\frac{\sigma \lbrack
(\gamma +\delta )b_{1}+2\epsilon b_{2}]q-2\xi }{\alpha q^{2}+\gamma
q+\epsilon }B.  \label{log}
\end{equation}%
Taking into account (\ref{quk}) and the constraints $\tilde{c}_{2}=\alpha ,%
\tilde{c}_{1}=\gamma ,\tilde{c}_{0}=\epsilon $, a comparison (\ref{rel})
with (\ref{second}) implies $q=W^{\prime }(p)$, i.e. 
\begin{equation*}
W^{\prime }=\frac{\delta \bar{W}^{\prime }-\epsilon b_{1}}{(\gamma +\delta
)b_{1}+2\epsilon b_{2}}.
\end{equation*}%
Differentiating (\ref{z}) and eliminating $\bar{W}^{\prime }$ from the above
relationship, finally, (\ref{sest}) reduces to

\begin{equation}
p_{t^{1}}=\partial _{x}(p(\delta f^{\prime }+\epsilon b_{1})+((\gamma
+\delta )b_{1}+2\epsilon b_{2})W),  \label{cor}
\end{equation}%
where all constants $b_{k}$ satisfy (\ref{rek}). Theorem is proved.

Equation (\ref{rel}) can be integrated in the parametric form%
\begin{equation}
p=\exp \int \frac{\delta dq}{\alpha q^{2}+\gamma q+\epsilon },\text{ \ \ \ }%
W=\int qdp  \label{ex}
\end{equation}%
Then all conservation law densities $h_{k}$ can be found by a substitution
of the inverse function $q(\lambda ,\mathbf{A})$ (expanded in the B\"{u}%
rmann--Lagrange series, see, for instance, \cite{ls}) in (\ref{ex}) at the
vicinity $q\rightarrow \infty $.

\subsection{Negative conservation laws and commuting flows}

If hydrodynamic chain (\ref{pyat}) is integrable, then an infinite series of
higher (positive) commuting flows (\ref{rek}) exist whose corresponding
Hamiltonian densities $h_{n}(A^{0},...,A^{n})$ depend on a finite set of
moments only.

If $\epsilon \neq 0$, hydrodynamic chain (\ref{pyat}) does not possess
negative local conservation laws, i.e. any negative conservation law density 
$h_{-n}$ depends on all negative moments $A^{-k}$. However, (\ref{rel}) is
reducible to the desirable form%
\begin{equation*}
\delta p\tilde{W}^{\prime \prime }=\alpha \tilde{W}^{\prime ^{2}}+(2\alpha
c+\gamma )\tilde{W}^{\prime }
\end{equation*}%
due to the shift $W=\tilde{W}+cp$, where the shift constant $c$ is a
solution of the quadratic equation $\alpha c^{2}+\gamma c+\epsilon =0$ (cf. (%
\ref{delta})). Then reduced Dorfman Poisson bracket (\ref{poj}) transforms
to the more simple form (expressed via new moments $\tilde{A}^{k}$)%
\begin{equation*}
\{\tilde{A}^{k}(x),\tilde{A}^{n}(x^{\prime })\}=[\tilde{\Gamma}^{kn}(\mathbf{%
\tilde{A}})D_{x}+D_{x}\tilde{\Gamma}^{nk}(\mathbf{\tilde{A}})]\delta
(x-x^{\prime }),\text{ \ \ \ }k,n=0,1,2,...
\end{equation*}%
where%
\begin{equation*}
\tilde{\Gamma}^{00}=\delta \tilde{A}^{0}+\xi ,\text{ \ }\tilde{\Gamma}%
^{kn}=\alpha k\tilde{A}^{k+n+1}+[(\gamma +2\alpha c)k+\delta ]\tilde{A}%
^{k+n},\text{ }k+n>0,
\end{equation*}%
and $A^{0}=\tilde{A}^{0},A^{1}=\tilde{A}^{1}+c\tilde{A}^{0},A^{2}=\tilde{A}%
^{2}+2c\tilde{A}^{1}+c^{2}\tilde{A}^{0},A^{3}=\tilde{A}^{3}+3c\tilde{A}%
^{2}+3c^{2}\tilde{A}^{1}+c^{3}\tilde{A}^{0}$ etc. This is a consequence
following from comparison (see (\ref{ak}))%
\begin{equation*}
dA^{k}=\delta \sum \epsilon _{m}a^{m}(W^{\prime }(a^{m}))^{k}da^{m}=\delta
\sum \epsilon _{m}a^{m}(\tilde{W}^{\prime }(a^{m})+c)^{k}da^{m}
\end{equation*}%
with%
\begin{equation*}
d\tilde{A}^{k}=\delta \sum \epsilon _{m}a^{m}(\tilde{W}^{\prime
}(a^{m}))^{k}da^{m}.
\end{equation*}

Thus, without loss of generality, the choice $\epsilon =0$ allows to
simplify all computations in this Section. In such a case, integrable
hydrodynamic chain (\ref{pyat})%
\begin{equation}
A_{t^{1}}^{0}=[2(\xi +\delta A^{0})f^{\prime \prime }+\delta f^{\prime
}]A_{x}^{0}+(\gamma +\delta )b_{1}A_{x}^{1},  \label{int}
\end{equation}%
\begin{equation*}
A_{t^{1}}^{k}=[\alpha kA^{k+1}+(\gamma k+2\delta )A^{k}]f^{\prime \prime
}A_{x}^{0}+\delta f^{\prime }A_{x}^{k}+(\gamma +\delta )b_{1}A_{x}^{k+1},%
\text{ \ }k=\pm 1,\pm 2,...
\end{equation*}%
possesses also an infinite series of lower (negative) local conservation
laws. For instance, the first negative conservation law is given by%
\begin{equation*}
\partial _{t^{1}}h_{-1}(A^{-1})=\delta \partial _{x}[f^{\prime
}(A^{0})h_{-1}(A^{-1})],
\end{equation*}%
where%
\begin{equation*}
(\ln h_{-1})^{\prime }=\frac{\delta }{(2\delta -\gamma )A^{-1}+\sigma
(\gamma +\delta )b_{1}},
\end{equation*}%
while (see (\ref{fun}))%
\begin{equation*}
f^{\prime \prime }(A^{0})=\frac{(\gamma +\delta )b_{1}}{\alpha A^{0}+\sigma
(\gamma +\delta )b_{1}}.
\end{equation*}%
Thus, the first negative commuting flow%
\begin{equation*}
A_{t^{-1}}^{k}=[\alpha (k-1)A^{k}+(\gamma (k-1)+2\delta
)A^{k-1}](h_{-1}^{\prime })_{x}+h_{-1}^{\prime }[-\alpha A_{x}^{k}+(\delta
-\gamma )A_{x}^{k-1}],\text{ \ }k=0,\pm 1,\pm 2,...
\end{equation*}%
is determined by the first negative Hamiltonian density $h_{-1}(A^{-1})$.
Its first non-negative\ conservation law (see (\ref{ham})) is%
\begin{eqnarray*}
\partial _{t^{-1}}h_{\ast } &=&\partial _{x}[b_{1}(\gamma +\delta
)A^{0}h_{-1}^{\prime }-\alpha A^{0}f^{\prime }h_{-1}^{\prime }-\delta
f^{\prime }h_{-1}+(2\delta -\gamma )f^{\prime }A^{-1}h_{-1}^{\prime }] \\
&& \\
&&+\partial _{x}\left[ h_{-1}^{\prime }\left( (\delta -\gamma )\underset{n=1}%
{\overset{\infty }{\sum }}b_{n}A^{n-1}-\alpha \underset{n=1}{\overset{\infty 
}{\sum }}b_{n}A^{n}\right) \right] ,
\end{eqnarray*}%
where constants $b_{n}$ are given by the reduced relationships (see (\ref%
{rek}))%
\begin{equation*}
\alpha nb_{n}+(\gamma (n+1)+\delta )b_{n+1}=0,\ n=1,2,...
\end{equation*}%
Under moment decomposition (\ref{ak}) the above hydrodynamic chain
transforms to the hydrodynamic reduction%
\begin{equation*}
a_{t^{-1}}^{i}=\delta \partial _{x}\left( \frac{a^{i}}{W^{\prime }(a^{i})}%
h_{-1}^{\prime }\right) .
\end{equation*}%
Thus, the generating function of conservation laws for the first negative
commuting flow%
\begin{equation}
p_{t^{-1}}=\delta \partial _{x}\left( \frac{p}{W^{\prime }(p)}h_{-1}^{\prime
}\right)   \label{anc}
\end{equation}%
is a particular case of (\ref{p}). Indeed, if $\epsilon =0$, then ordinary
differential equations (\ref{eq}) and (\ref{rel}) are connected by the
transformation%
\begin{equation*}
V(p)W^{\prime }(p)=p,
\end{equation*}%
where (see (\ref{exp}) and (\ref{ex}); $b$ is an integration constant and $%
b\neq 0$)%
\begin{equation}
V(p)=\frac{p}{\gamma }\left( bp^{-\frac{\gamma }{\delta }}-\alpha \right) ,%
\text{ \ \ }W(p)=\gamma \int \frac{dp}{bp^{-\frac{\gamma }{\delta }}-\alpha }%
,  \label{sol}
\end{equation}%
and (\ref{eq}) reduces to the particular case%
\begin{equation*}
VV^{\prime \prime }=\frac{\gamma }{\gamma -\delta }\left( V^{\prime }+\frac{%
\alpha }{\gamma }\right) \left( V^{\prime }+\frac{\alpha }{\delta }\right) .
\end{equation*}

While a computation of positive conservation laws is based on expansion (\ref%
{rim}) at the vicinity $q\rightarrow \infty $, negative conservation laws
can be found utilizing expansion (\ref{t}) at the vicinity $q\rightarrow 0$.
A generating function of conservation law densities%
\begin{equation*}
p=\left( \alpha +\frac{\gamma }{q(\lambda ,\mathbf{A})}\right) ^{-\delta
/\gamma }
\end{equation*}%
is an inverse expression to $q=W^{\prime }(p)$ (see (\ref{sol}); without
loss of generality, constant $b$ can be fixed to the unity). The first
series of higher (positive) conservation law densities can be found by a
substitution of an inverse expansion $q(\lambda ,\mathbf{A})$ from (\ref{rim}%
), where (see (\ref{log}))%
\begin{equation*}
(\ln B)^{\prime }=\frac{\alpha q-2\delta }{q(\alpha q+\gamma )},\text{ \ \ }%
C^{\prime }=\frac{\sigma (\gamma +\delta )b_{1}q-2\xi }{q(\alpha q+\gamma )}%
B,\text{ \ }q\rightarrow \infty .
\end{equation*}%
The second series of lower (negative) conservation law densities can be
found by a substitution of an inverse expansion $q(\lambda ,\mathbf{A})$
from (\ref{t}), where%
\begin{equation}
(\ln \tilde{B})^{\prime }=\frac{2\alpha q+\gamma -2\delta }{q(\alpha
q+\gamma )},\text{ \ \ }\tilde{C}^{\prime }=-\frac{\sigma (\gamma +\delta
)b_{1}}{(\alpha q+\gamma )q}\tilde{B},\text{ \ }q\rightarrow 0.  \label{cof}
\end{equation}%
Then substitution of these series of conservation law densities to the
generating function of conservation laws (\ref{cor})%
\begin{equation}
p_{t^{1}}=\partial _{x}(\delta f^{\prime }p+(\gamma +\delta )b_{1}W)
\label{genn}
\end{equation}%
allows to extract both series of conservation laws.

\textbf{Remark}: Functions $\tilde{B}(q)$ and $\tilde{C}(q)$ are found by
the same computation as in the previous Section. The Vlasov type kinetic
equation%
\begin{equation*}
\lambda _{t^{1}}=[\delta f^{\prime }+(\gamma +\delta )b_{1}W^{\prime
}]\lambda _{x}-\delta p\lambda _{p}f^{\prime \prime }A_{x}^{0}
\end{equation*}%
is connected with (\ref{genn}) by a semi-hodograph transformation $%
p(x,t;\lambda )\rightarrow \lambda (x,t;p)$. Taking into account the
negative part of integrable hydrodynamic chain (\ref{int})%
\begin{equation*}
A_{t^{1}}^{k}=[\alpha kA^{k+1}+(\gamma k+2\delta )A^{k}]f^{\prime \prime
}A_{x}^{0}+\delta f^{\prime }A_{x}^{k}+(\gamma +\delta )b_{1}A_{x}^{k+1},%
\text{ \ }k=-1,-2,...,
\end{equation*}%
a substitution of expansion (\ref{t}) into this Vlasov type kinetic equation
leads to (\ref{cof}).

\subsection{Generating function of commuting flows and conservation laws}

An integrable hierarchy of commuting hydrodynamic chains described in this
Section can be embedded into a sole generating function of conservation laws
and commuting flows (\ref{glob}). The function $G(p(\lambda ),p(\zeta ))$
can be found from the compatibility condition of (\ref{glob}) and (\ref{genn}%
)%
\begin{equation*}
\partial _{t^{1}}p(\lambda )=\partial _{x}[up(\lambda )+W(p(\lambda ))],%
\text{ \ \ \ }\partial _{\tau (\zeta )}p(\lambda )=\partial _{x}G(p(\lambda
),p(\zeta )).
\end{equation*}

\textbf{Theorem}: \textit{The function} $G(p(\lambda ),p(\zeta ))$ \textit{%
is defined by the quadrature}%
\begin{equation*}
dG(p(\lambda ),p(\zeta ))=\left( Q(p(\zeta ))-\frac{p(\zeta )R(p(\zeta ))}{%
W^{\prime }(p(\zeta ))-W^{\prime }(p(\lambda ))}\right) dp(\lambda )+\frac{%
p(\lambda )R(p(\zeta ))}{W^{\prime }(p(\zeta ))-W^{\prime }(p(\lambda ))}%
dp(\zeta ),
\end{equation*}%
\textit{where}%
\begin{equation}
R(p(\zeta ))=\frac{\delta W^{\prime \prime }(p(\zeta ))}{p(\zeta )},\text{ \
\ \ }Q^{\prime }(p(\zeta ))=\frac{\alpha }{\delta }R(p(\zeta )).  \label{zyk}
\end{equation}

\textbf{Proof}: The compatibility condition $\partial _{t^{1}}(\partial
_{\tau (\zeta )}p(\lambda ))=\partial _{\tau (\zeta )}(\partial
_{t^{1}}p(\lambda ))$ implies%
\begin{equation}
\partial _{\tau (\zeta )}u=Q(p(\zeta ))u_{x}+R(p(\zeta ))\partial
_{x}p(\zeta ),  \label{zik}
\end{equation}%
\begin{equation*}
\frac{\partial G(p(\lambda ),p(\zeta ))}{\partial p(\lambda )}=Q(p(\zeta ))-%
\frac{p(\zeta )R(p(\zeta ))}{W^{\prime }(p(\zeta ))-W^{\prime }(p(\lambda ))}%
,\text{ \ \ }\frac{\partial G(p(\lambda ),p(\zeta ))}{\partial p(\zeta )}=%
\frac{p(\lambda )R(p(\zeta ))}{W^{\prime }(p(\zeta ))-W^{\prime }(p(\lambda
))},
\end{equation*}%
where functions $Q(p(\zeta ))$ and $R(p(\zeta ))$ are not yet determined.
The compatibility condition%
\begin{equation*}
\frac{\partial }{\partial p(\lambda )}\frac{\partial G(p(\lambda ),p(\zeta ))%
}{\partial p(\zeta )}=\frac{\partial }{\partial p(\zeta )}\frac{\partial
G(p(\lambda ),p(\zeta ))}{\partial p(\lambda )}
\end{equation*}%
leads to (\ref{zyk}). Theorem is proved.

Moreover (\ref{zik}) reduces to (see the second equation in (\ref{zyk}))%
\begin{equation*}
\partial _{\tau (\zeta )}e^{\alpha u/\delta }=(Q(p(\zeta ))e^{\alpha
u/\delta })_{x}
\end{equation*}%
It is easy to see the conservation law density $e^{\alpha u/\delta }$ is
nothing but a momentum density $A^{0}$ (up to unessential additive and
multiplicative constants).

\subsection{Integrable three dimensional quasilinear equations of the second
order}

Compatibility conditions $(p_{t^{k}})_{t^{n}}=(p_{t^{n}})_{t^{k}}$ lead to
integrable three dimensional hydrodynamic type systems. If $k=1$ and $n=-1$,
then three dimensional hydrodynamic type system%
\begin{equation}
u_{t^{-1}}+v_{t^{0}}+\frac{\alpha }{\delta }vu_{t^{0}}=0,\text{ \ \ }%
v_{t^{1}}=uv_{t^{0}}+\frac{\gamma -\delta }{\delta }vu_{t^{0}}  \label{sys}
\end{equation}%
is determined by the dispersionless Lax pair (see (\ref{anc}) and (\ref{genn}%
))%
\begin{equation}
p_{t^{1}}=\partial _{t^{0}}(up+W(p)),\text{ \ \ }p_{t^{-1}}=\partial
_{t^{0}}\left( \frac{p}{W^{\prime }(p)}v\right) ,  \label{bark}
\end{equation}%
where%
\begin{equation*}
u=\frac{\delta }{(\gamma +\delta )b_{1}}f^{\prime },\text{ \ \ }v=\delta
h_{-1}^{\prime }.
\end{equation*}%
Three dimensional hydrodynamic type system (\ref{sys}) can be written in the
conservative form ($\gamma \neq \delta $)%
\begin{equation*}
(e^{\alpha u/\delta })_{t^{-1}}+\frac{\alpha }{\delta }(ve^{\alpha u/\delta
})_{t^{0}}=0,\text{ \ \ }(v^{\frac{\delta }{\gamma -\delta }})_{t^{1}}=(uv^{%
\frac{\delta }{\gamma -\delta }})_{t^{0}}.
\end{equation*}%
Introducing the potential function $z$ such that (see the first conservation
law)%
\begin{equation*}
u=\frac{\delta }{\alpha }\ln z_{t^{0}},\text{ \ \ }v=-\frac{\delta z_{t^{-1}}%
}{\alpha z_{t^{0}}},
\end{equation*}%
(\ref{sys}) reduces to the three dimensional quasilinear equation of the
second order 
\begin{equation*}
\alpha (z_{t^{0}}z_{t^{-1}t^{1}}-z_{t^{-1}}z_{t^{0}t^{1}})=\delta
z_{t^{0}}\ln z_{t^{0}}\cdot z_{t^{-1}t^{0}}+(\gamma -\delta -\delta \ln
z_{t^{0}})z_{t^{-1}}z_{t^{0}t^{0}};
\end{equation*}%
introducing the potential function $\tilde{z}$ such that (see the second
conservation law)%
\begin{equation*}
v=(\tilde{z}_{t^{0}})^{\frac{\gamma -\delta }{\delta }},\text{ \ \ }u=\frac{%
\tilde{z}_{t^{1}}}{\tilde{z}_{t^{0}}},
\end{equation*}%
(\ref{sys}) reduces to another three dimensional quasilinear equation of the
second order%
\begin{equation*}
\delta (\tilde{z}_{t^{0}}\tilde{z}_{t^{1}t^{-1}}-\tilde{z}_{t^{1}}\tilde{z}%
_{t^{0}t^{-1}})+(\tilde{z}_{t^{0}})^{\frac{\gamma -\delta }{\delta }%
}[(\gamma -\delta )\tilde{z}_{t^{0}}-\alpha \tilde{z}_{t^{1}}]\tilde{z}%
_{t^{0}t^{0}}+\alpha (\tilde{z}_{t^{0}})^{\frac{\gamma -\delta }{\delta }+1}%
\tilde{z}_{t^{0}t^{1}}=0.
\end{equation*}%
A substitution (where the potential function $\psi $ is introduced for (\ref%
{bark}))%
\begin{equation*}
u=\frac{\psi _{t^{1}}-W(\psi _{t^{0}})}{\psi _{t^{0}}},\text{ \ \ \ }v=\frac{%
W^{\prime }(\psi _{t^{0}})}{\psi _{t^{0}}}\psi _{t^{-1}}
\end{equation*}%
into (\ref{sys}) implies the integrable quasilinear three dimensional
equation of the second order%
\begin{equation*}
\psi _{t^{-1}t^{1}}+\frac{\alpha W^{\prime }(\psi _{t^{0}})}{\delta \psi
_{t^{0}}}\psi _{t^{-1}}\psi _{t^{0}t^{1}}-\frac{\psi _{t^{1}}-W(\psi
_{t^{0}})}{\psi _{t^{0}}}\psi _{t^{0}t^{-1}}
\end{equation*}%
\begin{equation*}
+\frac{W^{\prime }(\psi _{t^{0}})}{\delta \psi _{t^{0}}^{2}}\left( \alpha
W(\psi _{t^{0}})+(\gamma -\delta )\psi _{t^{0}}-\alpha \psi _{t^{1}}\right)
\psi _{t^{-1}}\psi _{t^{0}t^{0}}=0.
\end{equation*}

If $k=1$ and $n=2$, then the dispersionless Lax pair (see (\ref{gen}))%
\begin{equation}
p_{t^{1}}=\partial _{t^{0}}(up+W(p)),\text{ \ \ }p_{t^{2}}=\partial
_{t^{0}}(wp+spW^{\prime }(p))  \label{murk}
\end{equation}%
determines the constraint $s=e^{\alpha u/\delta }$ and the three dimensional
hydrodynamic type system%
\begin{equation}
(e^{\alpha u/\delta })_{t^{1}}=\left( w+\left( u-\frac{2\delta +\gamma }{%
\alpha }\right) e^{\alpha u/\delta }\right) _{t^{0}},\text{ \ \ }%
u_{t^{2}}+uw_{t^{0}}=w_{t^{1}}+wu_{t^{0}},  \label{ro}
\end{equation}%
which is equivalent to the three dimensional quasilinear equation of the
second order%
\begin{equation*}
\alpha \delta z_{t^{0}t^{2}}=\alpha ^{2}z_{t^{0}}z_{t^{1}t^{1}}+\alpha
(\gamma +\delta -2\delta \ln z_{t^{0}})z_{t^{0}}z_{t^{0}t^{1}}+\delta
\lbrack \delta z_{t^{0}}(\ln z_{t^{0}})^{2}-(\gamma +2\delta )z_{t^{0}}(\ln
z_{t^{0}}-1)+\alpha z_{t^{1}}]z_{t^{0}t^{0}},
\end{equation*}%
where $z$ is a potential function for the conservation law in (\ref{ro}). A
substitution (where the potential function $\psi $ is introduced for (\ref%
{murk}))%
\begin{equation*}
u=\frac{\psi _{t^{1}}-W(\psi _{t^{0}})}{\psi _{t^{0}}},\text{ \ \ \ }w=\frac{%
\psi _{t^{2}}}{\psi _{t^{0}}}-W^{\prime }(\psi _{t^{0}})\exp \left( \frac{%
\alpha \psi _{t^{1}}-\alpha W(\psi _{t^{0}})}{\delta \psi _{t^{0}}}\right) 
\end{equation*}%
into (\ref{ro}) implies the integrable quasilinear three dimensional
equation of the second order%
\begin{equation*}
\delta \exp \left( \frac{\alpha W(\psi _{t^{0}})-\alpha \psi _{t^{1}}}{%
\delta \psi _{t^{0}}}\right) \cdot \left( \psi _{t^{0}t^{2}}-\frac{\psi
_{t^{2}}}{\psi _{t^{0}}}\psi _{t^{0}t^{0}}\right) =\alpha \psi
_{t^{1}t^{1}}+\left( \gamma +\delta -2\alpha \frac{\psi _{t^{1}}-W(\psi
_{t^{0}})}{\psi _{t^{0}}}\right) \psi _{t^{0}t^{1}}
\end{equation*}%
\begin{equation*}
+\left( \alpha \frac{\lbrack \psi _{t^{1}}-W(\psi _{t^{0}})]^{2}}{\psi
_{t^{0}}^{2}}-(\gamma +\delta )\frac{\psi _{t^{1}}-W(\psi _{t^{0}})}{\psi
_{t^{0}}}-\delta W^{\prime }(\psi _{t^{0}})\right) \psi _{t^{0}t^{0}}.
\end{equation*}

The next integrable three dimensional quasilinear equation of the second
order follows from (\ref{genn}) and (\ref{zik})%
\begin{equation*}
\partial _{\tau }e^{\alpha u/\delta }=\partial _{t^{0}}(Q(p)e^{\alpha
u/\delta }),\text{ \ \ }\partial _{t^{1}}p=\partial _{t^{0}}(up+W(p)).
\end{equation*}%
Introducing a potential function $z$ such that (utilizing the first
conservation law) $z_{t^{0}}=e^{\alpha u/\delta },z_{\tau }=Q(p)z_{t^{0}}$,
the second conservation law transforms to%
\begin{equation*}
z_{t^{0}}z_{\tau t^{1}}-z_{\tau }z_{t^{0}t^{1}}=\frac{\delta p(s)}{\alpha
p^{\prime }(s)}z_{t^{0}}z_{t^{0}t^{0}}+(\frac{\delta }{\alpha }\ln
z_{t^{0}}+W^{\prime }(p(s)))(z_{t^{0}}z_{\tau t^{0}}-z_{\tau
}z_{t^{0}t^{0}}),
\end{equation*}%
where $p(s)$ is an inverse function to $Q(p)$ and $s=z_{\tau }/z_{t^{0}}$.
Introducing a potential function $\psi $ such that (utilizing the second
conservation law) $\psi _{t^{0}}=p,\psi _{t^{1}}=up+W(p)$, the first
conservation law transforms to another integrable three dimensional
quasilinear equation of the second order%
\begin{equation*}
\psi _{t^{0}}\psi _{t^{1}\tau }=(\psi _{t^{1}}\psi _{t^{0}\tau }+\psi
_{t^{0}}W^{\prime }(\psi _{t^{0}})-W(\psi _{t^{0}}))\psi _{t^{0}\tau }+\psi
_{t^{0}}Q(\psi _{t^{0}})\psi _{t^{0}t^{1}}
\end{equation*}%
\begin{equation*}
+\left( Q(\psi _{t^{0}})(W(\psi _{t^{0}})-\psi _{t^{0}}W^{\prime }(\psi
_{t^{0}})-\psi _{t^{1}})+\frac{\delta }{\alpha }\psi _{t^{0}}^{2}Q^{\prime
}(\psi _{t^{0}})\right) \psi _{t^{0}t^{0}}.
\end{equation*}

\section{The second degenerate level $\protect\beta =0$ and $\protect\delta %
=0$}

In Section \textbf{3}, the Theorem was formulated for the case $\beta =0$
but $\delta \neq 0$. This Section is devoted to the second degeneration $%
\delta =0$.

\textbf{Theorem}: \textit{Suppose }$\beta =0$ \textit{and} $\delta =0$,%
\textit{\ in such a case, Dorfman Poisson bracket} (\ref{poi}) \textit{%
reduces to }(\ref{pou}) \textit{under the moment decomposition}%
\begin{equation}
dA^{k}=\sum \epsilon _{m}(U^{\prime }(a^{m}))^{k}da^{m},\text{ \ \ \ }%
k=0,1,...,  \label{sub}
\end{equation}%
\textit{where the function} $U(p)$ \textit{satisfies the ODE}%
\begin{equation}
U^{\prime \prime }=\alpha U^{\prime ^{2}}+\gamma U^{\prime }+\epsilon ,
\label{rem}
\end{equation}%
\textit{and }$\xi $\textit{\ is an integration constant of the simplest
constraint}%
\begin{equation}
\sum \epsilon _{m}=\xi .  \label{ost}
\end{equation}

\textbf{Proof}: Under the substitution $V=\beta U+\delta p+1$, (\ref{sec})
reduces to the form (cf. (\ref{rel}))%
\begin{equation*}
\beta UU^{\prime \prime }+(\delta p+1)U^{\prime \prime }=\alpha U^{\prime
^{2}}+\gamma U^{\prime }+\epsilon 
\end{equation*}%
Thus, the double degenerate case $\beta =0$ and then $\delta =0$ is
associated with (\ref{rem}). A substitution of moment decomposition (\ref%
{dec}) into the Dorfman Poisson bracket (cf. (\ref{poi}))%
\begin{equation}
\{A^{k}(x),A^{n}(x^{\prime })\}=[\Gamma ^{kn}(\mathbf{A})D_{x}+D_{x}\Gamma
^{nk}(\mathbf{A})]\delta (x-x^{\prime }),\text{ \ \ \ }k,n=0,1,2,...
\label{pog}
\end{equation}%
where ($\alpha ,\gamma ,\epsilon ,\xi $ are arbitrary constants)%
\begin{equation*}
\Gamma ^{00}=\xi ,\text{ \ }\Gamma ^{kn}=k(\alpha A^{k+n+1}+\gamma
A^{k+n}+\epsilon A^{k+n-1}),\text{ }k+n>0,
\end{equation*}%
implies the recursive relationships%
\begin{equation*}
n(\alpha U_{k+n+1}+\gamma U_{k+n}+\epsilon U_{k+n-1})=U_{k}U_{n}^{\prime },%
\text{ \ \ }k,n=0,1,2,...,
\end{equation*}%
where $U_{n}(a^{i})=f_{n,i}^{\prime }/\epsilon _{i}$. Following the general
case (see (\ref{mom}), (\ref{sec}) and below), suppose that a solution of
this system is given by $U_{m}=U_{0}(U^{\prime })^{m}$, where $U_{0}(p)$ and 
$U(p)$ are not yet determined. If $n=0$, then $U_{0}(p)=1$ (an arbitrary
nonzero additive constant is fixed to the unity here). A substitution of the
above ansatz into the recursive relationships leads to (\ref{rem}).
Moreover, taking into account (\ref{rem}), an integration of the
differential $d(\alpha A^{k+2}+\gamma A^{k+1}+\epsilon A^{k})$ leads (see (%
\ref{sub})) to an infinite series of constraints%
\begin{equation*}
k(\alpha A^{k+1}+\gamma A^{k}+\epsilon A^{k-1})+\xi _{k}=\sum \epsilon
_{m}(U^{\prime }(a^{m}))^{k},\text{ \ }k=0,1,...,
\end{equation*}%
where $\xi _{k}$ are integration constants. It is easy to see, if $k=0$,
then $\xi _{0}=\xi $ (see (\ref{ost})). The Theorem is proved.

For any positive integer $M$, an arbitrary Hamiltonian density $%
h_{M}(A^{0},...,A^{M})$ and Dorfman Poisson bracket (\ref{pog}) determine a
hydrodynamic chain (see the previous Sections), whose Hamiltonian
hydrodynamic reduction (\ref{sisa}) is presented in the symmetric form (see (%
\ref{sub}) and \cite{algebra}; here $h_{M,m}\equiv \partial h/\partial
A^{m},m=0,1,...,M$)%
\begin{equation*}
a_{t}^{i}=\frac{1}{\epsilon _{i}}\partial _{x}\left( \overset{M}{\underset{%
m=0}{\sum }}h_{M,m}\frac{\partial A^{m}}{\partial a^{i}}\right) \equiv
\partial _{x}\left( \overset{M}{\underset{m=0}{\sum }}(U^{\prime
}(a^{i}))^{m}h_{M,m}\right) .
\end{equation*}%
Then such an integrable hydrodynamic chain possesses the generating function
of conservation laws (cf. (\ref{gena}))%
\begin{equation}
p_{t}=\partial _{x}\left( \overset{M}{\underset{m=0}{\sum }}(U^{\prime
}(p))^{m}h_{M,m}\right) .  \label{gem}
\end{equation}%
In this case, the Hamiltonian density satisfies some overdetermined system.
Since this double degenerate case ($\beta =0$ and $\delta =0$) is very
familiar to the degenerate case ($\beta =0$), very similar results are
presented below.

\textbf{Lemma}: \textit{The hydrodynamic chain}%
\begin{equation}
A_{t^{1}}^{0}=\left( 2\xi f^{\prime }+\epsilon b_{1}A^{0}+(\gamma
b_{1}+2\epsilon b_{2})A^{1}\right) _{x},  \label{sem}
\end{equation}%
\begin{equation*}
A_{t^{1}}^{k}=k(\alpha A^{k+1}+\gamma A^{k}+\epsilon A^{k-1})f^{\prime
\prime }A_{x}^{0}+\epsilon b_{1}A_{x}^{k}+(\gamma b_{1}+2\epsilon
b_{2})A_{x}^{k+1},\text{ \ }k=1,2,...
\end{equation*}%
\textit{is determined by Hamiltonian density }(\ref{ham}) \textit{where} $%
b_{1}$ \textit{and} $b_{2}$ \textit{are arbitrary constants, while all other
constants} $b_{n}$ \textit{satisfy the recursive relationships}%
\begin{equation}
\alpha (n-1)b_{n-1}+\gamma nb_{n}+\epsilon (n+1)b_{n+1}=0,\text{ \ }n=2,3,...
\label{rec}
\end{equation}

\textbf{Proof}: can obtained by a straightforward calculation.

Moreover, integrable hydrodynamic chain (\ref{sem}) is associated with the
generating function of conservation laws (see (\ref{gem}))%
\begin{equation}
p_{t^{1}}=\partial _{x}(f^{\prime }(A^{0})+\bar{U}(p)),  \label{third}
\end{equation}%
where%
\begin{equation}
\bar{U}(p)=\overset{\infty }{\underset{m=1}{\sum }}b_{m}(U^{\prime }(p))^{m}.
\label{zuk}
\end{equation}%
In this case (see \cite{MaksGen}),%
\begin{equation}
\bar{U}^{\prime \prime }=c_{2}\bar{U}^{\prime ^{2}}+c_{1}\bar{U}^{\prime
}+c_{0},  \label{zz}
\end{equation}%
where $c_{k}$ are some constants.

\textbf{Theorem}: \textit{Integrable hydrodynamic chain} (\ref{sem}) \textit{%
is determined by Hamiltonian density} (\ref{ham}), \textit{where}%
\begin{equation}
f^{\prime \prime }(A^{0})=\frac{\gamma b_{1}+2\epsilon b_{2}}{\sigma +\alpha
A^{0}},  \label{hami}
\end{equation}%
$\sigma $ \textit{is an integration constant and} $b_{k}$ \textit{satisfy} (%
\ref{rec}).

\textbf{Proof}: Under the semi-hodograph transformation $p(x,t;\lambda
)\rightarrow \lambda (x,t;p)$, (\ref{third}) reduces to the Vlasov type
kinetic equation (see \cite{OPS})%
\begin{equation*}
\lambda _{t^{1}}=\bar{U}^{\prime }(p)\lambda _{x}-\lambda _{p}f^{\prime
\prime }(A^{0})A_{x}^{0}.
\end{equation*}%
A substitution $\lambda (p,\mathbf{A})$ leads to%
\begin{equation}
\lambda _{p}=\left( \frac{\bar{U}^{\prime }(p)-\epsilon b_{1}}{f^{\prime
\prime }(A^{0})}-2\xi -\overset{\infty }{\underset{k=1}{\sum }}\frac{%
k(\alpha A^{k+1}+\gamma A^{k}+\epsilon A^{k-1})}{q^{k}}\right) \lambda _{0}
\label{con}
\end{equation}%
and (\ref{d}), where (instead (\ref{q}) and (\ref{quk}))%
\begin{equation}
q=\frac{\bar{U}^{\prime }(p)-\epsilon b_{1}}{\gamma b_{1}+2\epsilon b_{2}}.
\label{quq}
\end{equation}%
Taking into account (see (\ref{zz}) and (\ref{quq}))%
\begin{equation}
\frac{\bar{U}^{\prime \prime }}{\gamma b_{1}+2\epsilon b_{2}}=\tilde{c}%
_{2}q^{2}+\tilde{c}_{1}q+\tilde{c}_{0},  \label{short}
\end{equation}%
a substitution (\ref{rim}) into (\ref{con}) implies the constraints $\tilde{c%
}_{2}=\alpha ,\tilde{c}_{1}=\gamma ,\tilde{c}_{0}=\epsilon $, (\ref{hami})
and%
\begin{equation}
(\ln B)^{\prime }=\frac{\alpha q^{2}-\epsilon }{q(\alpha q^{2}+\gamma
q+\epsilon )},\text{ \ \ }C^{\prime }=\frac{\sigma q-2\xi }{\alpha
q^{2}+\gamma q+\epsilon }B.  \label{ln}
\end{equation}%
Taking into account (\ref{quq}) and the constraints $\tilde{c}_{2}=\alpha ,%
\tilde{c}_{1}=\gamma ,\tilde{c}_{0}=\epsilon $, a comparison (\ref{rem})
with (\ref{short}) implies $q=U^{\prime }(p)$, i.e.%
\begin{equation*}
U^{\prime }=\frac{\bar{U}^{\prime }-\epsilon b_{1}}{\gamma b_{1}+2\epsilon
b_{2}}.
\end{equation*}%
Differentiating (\ref{zuk}) and eliminating $\bar{U}^{\prime }$ from the
above relationship, finally, (\ref{third}) reduces to%
\begin{equation}
p_{t^{1}}=\partial _{x}(f^{\prime }(A^{0})+(\gamma b_{1}+2\epsilon
b_{2})U(p)+\epsilon b_{1}p),  \label{cos}
\end{equation}%
where constants $b_{k}$ satisfy (\ref{rec}). Theorem is proved.

Equation (\ref{rem}) can be integrated in the parametric form%
\begin{equation}
p=\int \frac{dq}{\alpha q^{2}+\gamma q+\epsilon },\text{ \ \ }U=\int \frac{%
qdq}{\alpha q^{2}+\gamma q+\epsilon }.  \label{ext}
\end{equation}%
Then all conservation law densities $h_{k}$ can be found by a substitution
of the inverse function $q(\lambda ,\mathbf{A})$ (expanded in the B\"{u}%
rmann--Lagrange series, see, for instance, \cite{ls}) in (\ref{ext}) at the
vicinity $q\rightarrow \infty $.

\subsection{Negative conservation laws and commuting flows}

If hydrodynamic chain (\ref{sem}) is integrable, then an infinite series of
higher (positive) commuting flows exist whose corresponding Hamiltonian
densities $h_{n}(A^{0},...,A^{n})$ depend on a finite set of moments only.

If $\epsilon \neq 0$, hydrodynamic chain (\ref{sem}) does not possess
negative local conservation laws, i.e. any negative conservation law density 
$h_{-n}$ depends on all negative moments $A^{-k}$. However, (\ref{rem}) is
reducible to the desirable form%
\begin{equation*}
\tilde{U}^{\prime \prime }=\alpha \tilde{U}^{\prime ^{2}}+(\gamma +2\alpha c)%
\tilde{U}^{\prime }
\end{equation*}%
due to the shift $U=\tilde{U}+cp$, where the shift constant $c$ is a
solution of the quadratic equation $\alpha c^{2}+\gamma c+\epsilon =0$ (cf. (%
\ref{delta})). Then reduced Dorfman Poisson bracket (\ref{pog}) transforms
to the more simple form (expressed via new moments $\tilde{A}^{k}$)%
\begin{equation*}
\{\tilde{A}^{k}(x),\tilde{A}^{n}(x^{\prime })\}=[\tilde{\Gamma}^{kn}(\mathbf{%
\tilde{A}})D_{x}+D_{x}\tilde{\Gamma}^{nk}(\mathbf{\tilde{A}})]\delta
(x-x^{\prime }),\text{ \ \ \ }k,n=0,1,2,...
\end{equation*}%
where%
\begin{equation*}
\tilde{\Gamma}^{00}=\xi ,\text{ \ }\tilde{\Gamma}^{kn}=k(\alpha \tilde{A}%
^{k+n+1}+(\gamma +2\alpha c)\tilde{A}^{k+n}),\text{ }k+n>0,
\end{equation*}%
and $A^{0}=\tilde{A}^{0},A^{1}=\tilde{A}^{1}+c\tilde{A}^{0},A^{2}=\tilde{A}%
^{2}+2c\tilde{A}^{1}+c^{2}\tilde{A}^{0},A^{3}=\tilde{A}^{3}+3c\tilde{A}%
^{2}+3c^{2}\tilde{A}^{1}+c^{3}\tilde{A}^{0}$ etc. This is a consequence
following from comparison (see (\ref{sub}))%
\begin{equation*}
dA^{k}=\sum \epsilon _{m}(U^{\prime }(a^{m}))^{k}da^{m}=\sum \epsilon _{m}(%
\tilde{U}^{\prime }(a^{m})+c)^{k}da^{m}
\end{equation*}%
with%
\begin{equation*}
d\tilde{A}^{k}=\sum \epsilon _{m}(\tilde{U}^{\prime }(a^{m}))^{k}da^{m}.
\end{equation*}

Thus, without loss of generality, the choice $\epsilon =0$ allows to
simplify all computations in this Section. In such a case, integrable
hydrodynamic chain (\ref{sem})%
\begin{equation}
A_{t^{1}}^{0}=\left( 2\xi f^{\prime }+\gamma b_{1}A^{1}\right) _{x},\text{ \
\ }A_{t^{1}}^{k}=k(\alpha A^{k+1}+\gamma A^{k})f^{\prime \prime
}A_{x}^{0}+\gamma b_{1}A_{x}^{k+1},\text{ \ }k=\pm 1,\pm 2,...  \label{cin}
\end{equation}%
possesses also an infinite series of lower (negative) local conservation
laws. For instance, the first negative conservation law is given by ($\gamma
\neq 0$)%
\begin{equation}
\partial _{t^{1}}\ln (\gamma A^{-1}-\sigma )=-\frac{\gamma ^{2}b_{1}}{\alpha 
}\partial _{x}\ln (\alpha A^{0}+\sigma )  \label{cink}
\end{equation}%
where (see (\ref{hami}))%
\begin{equation*}
f^{\prime }(A^{0})=\frac{\gamma b_{1}}{\alpha }\ln (\alpha A^{0}+\sigma ).
\end{equation*}%
Thus, the first negative commuting flow%
\begin{equation*}
A_{t^{-1}}^{k}=(k-1)(\alpha A^{k}+\gamma A^{k-1})(h_{-1}^{\prime
})_{x}-h_{-1}^{\prime }(\alpha A_{x}^{k}+\gamma A_{x}^{k-1})
\end{equation*}%
is determined by the first negative Hamiltonian density $h_{-1}(A^{-1})=\ln
(\gamma A^{-1}-\sigma )$. Its first non-negative\ conservation law (see (\ref%
{ham})) is%
\begin{equation*}
\partial _{t^{-1}}h_{\ast }=\left( \frac{\gamma ^{2}}{\alpha }b_{1}\ln
(\sigma +\alpha A^{0})+\gamma b_{1}A^{0}h_{-1}^{\prime }-(\alpha
A^{0}+\gamma A^{-1})f^{\prime }h_{-1}^{\prime }-\underset{n=1}{\overset{%
\infty }{\sum }}b_{n}(\alpha A^{n}+\gamma A^{n-1})h_{-1}^{\prime }\right)
_{x}
\end{equation*}%
where constants $b_{n}$ are given by the reduced relationships (see (\ref%
{rec}))%
\begin{equation*}
\alpha nb_{n}+\gamma (n+1)b_{n+1}=0,\text{ \ }n=1,2,...
\end{equation*}%
Under moment decomposition (\ref{sub}) the above hydrodynamic chain
transforms to the hydrodynamic reduction%
\begin{equation*}
a_{t^{-1}}^{i}=\partial _{x}\frac{h_{-1}^{\prime }}{U^{\prime }(a^{i})}.
\end{equation*}%
Thus, the generating function of conservation laws for the first negative
commuting flow%
\begin{equation}
p_{t^{-1}}=\partial _{x}\frac{h_{-1}^{\prime }}{U^{\prime }(p)}.  \label{neg}
\end{equation}%
is a particular case of (\ref{p}). Indeed, if $\epsilon =0$ and $\delta =0$,
then ordinary differential equations (\ref{eq}) and (\ref{rem}) are
connected by the transformation%
\begin{equation*}
V(p)U^{\prime }(p)=1,
\end{equation*}%
where (see (\ref{exp}) and (\ref{ext}))%
\begin{equation}
V(p)=\frac{e^{-\gamma p}-\alpha }{\gamma },\text{ \ \ }U(p)=\frac{1}{\alpha }%
\ln \gamma -\frac{1}{\alpha }\ln (1-\alpha e^{\gamma p})  \label{rez}
\end{equation}%
and (\ref{eq}) reduces to the particular case%
\begin{equation*}
VV^{\prime \prime }=V^{\prime }\left( V^{\prime }+\alpha \right) .
\end{equation*}

While a computation of positive conservation laws is based on expansion (\ref%
{rim}) at the vicinity $q\rightarrow \infty $, negative conservation laws
can be found utilizing expansion (\ref{t}) at the vicinity $q\rightarrow 0$.
A generating function of conservation law densities%
\begin{equation*}
p=-\frac{1}{\gamma }\ln \left( \alpha +\frac{\gamma }{q(\lambda ,\mathbf{A})}%
\right) 
\end{equation*}%
is an inverse expression to $q=U^{\prime }(p)$ (see (\ref{rez})). The first
series of higher (positive) conservation law densities can be found by a
substitution of an inverse expansion $q(\lambda ,\mathbf{A})$ from (\ref{rim}%
) (see (\ref{ln}))%
\begin{equation*}
\lambda (q,\mathbf{A})=\sigma q-2\xi \ln q+(\alpha q+\gamma )\overset{\infty 
}{\underset{k=0}{\sum }}\frac{A^{k}}{q^{k}},\text{ \ }q\rightarrow \infty .
\end{equation*}%
The second series of lower (negative) conservation law densities can be
found by a substitution of an inverse expansion $q(\lambda ,\mathbf{A})$
from (\ref{t})%
\begin{equation}
\lambda (q,\mathbf{A})=(\alpha q^{2}+\gamma q)\overset{-1}{\underset{%
k=-\infty }{\sum }}q^{-k-1}A^{k}-\sigma q,\text{ \ }q\rightarrow 0.
\label{gemma}
\end{equation}%
Then substitution of these series of conservation law densities into
generating function of conservation laws (\ref{cos})%
\begin{equation}
p_{t^{1}}=\frac{\gamma b_{1}}{\alpha }\partial _{x}\ln \frac{\alpha
A^{0}+\sigma }{1-\alpha e^{\gamma p}}  \label{agen}
\end{equation}%
allows to extract both series of conservation laws (the above expression is
obtained by virtue of (\ref{rez})).

\textbf{Remark}: Expansion (\ref{gemma}) is found by the same computation as
in the previous Sections. The Vlasov type kinetic equation%
\begin{equation*}
\lambda _{t^{1}}=\gamma b_{1}\left( U^{\prime }(p)\lambda _{x}-\lambda _{p}%
\frac{A_{x}^{0}}{\alpha A^{0}+\sigma }\right) 
\end{equation*}%
is connected with (\ref{agen}) by a semi-hodograph transformation $%
p(x,t;\lambda )\rightarrow \lambda (x,t;p)$. Taking into account the
negative part of integrable hydrodynamic chain (\ref{cin})%
\begin{equation*}
A_{t^{1}}^{k}=k(\alpha A^{k+1}+\gamma A^{k})f^{\prime \prime
}A_{x}^{0}+\gamma b_{1}A_{x}^{k+1},\text{ \ }k=-1,-2,...,
\end{equation*}%
a substitution of expansion (\ref{t}) into this Vlasov type kinetic equation
leads to (\ref{gemma}).

\subsection{Generating function of commuting flows and conservation laws}

An integrable hierarchy of commuting hydrodynamic chains described in this
Section can be embedded into a sole generating function of conservation laws
and commuting flows (\ref{glob}). The function $G(p(\lambda ),p(\zeta ))$
can be found from the compatibility condition of (\ref{glob}) and (\ref{cos})%
\begin{equation*}
\partial _{t^{1}}p(\lambda )=\partial _{x}[u+U(p(\lambda ))],\text{ \ \ \ }%
\partial _{\tau (\zeta )}p(\lambda )=\partial _{x}G(p(\lambda ),p(\zeta )).
\end{equation*}

\textbf{Theorem}: \textit{The function} $G(p(\lambda ),p(\zeta ))$ \textit{%
is defined by the quadrature}%
\begin{equation*}
dG(p(\lambda ),p(\zeta ))=\left( Q(p(\zeta ))-\frac{R(p(\zeta ))}{U^{\prime
}(p(\zeta ))-U^{\prime }(p(\lambda ))}\right) dp(\lambda )+\frac{R(p(\zeta ))%
}{U^{\prime }(p(\zeta ))-U^{\prime }(p(\lambda ))}dp(\zeta ),
\end{equation*}%
\textit{where}%
\begin{equation}
R(p(\zeta ))=\exp [2\alpha U(p(\zeta ))+\gamma p(\zeta )],\text{ \ \ \ }%
Q^{\prime }(p(\zeta ))=\alpha R(p(\zeta ))  \label{smik}
\end{equation}

\textbf{Proof}: The compatibility condition $\partial _{t^{1}}(\partial
_{\tau (\zeta )}p(\lambda ))=\partial _{\tau (\zeta )}(\partial
_{t^{1}}p(\lambda ))$ implies%
\begin{equation}
\partial _{\tau (\zeta )}u=Q(p(\zeta ))u_{x}+R(p(\zeta ))\partial
_{x}p(\zeta ),  \label{smyk}
\end{equation}%
\begin{equation*}
\frac{\partial G(p(\lambda ),p(\zeta ))}{\partial p(\lambda )}=Q(p(\zeta ))-%
\frac{R(p(\zeta ))}{U^{\prime }(p(\zeta ))-U^{\prime }(p(\lambda ))},\text{
\ \ \ }\frac{\partial G(p(\lambda ),p(\zeta ))}{\partial p(\zeta )}=\frac{%
R(p(\zeta ))}{U^{\prime }(p(\zeta ))-U^{\prime }(p(\lambda ))},
\end{equation*}%
where functions $Q(p(\zeta ))$ and $R(p(\zeta ))$ are not yet determined.
The compatibility condition%
\begin{equation*}
\frac{\partial }{\partial p(\lambda )}\frac{\partial G(p(\lambda ),p(\zeta ))%
}{\partial p(\zeta )}=\frac{\partial }{\partial p(\zeta )}\frac{\partial
G(p(\lambda ),p(\zeta ))}{\partial p(\lambda )}
\end{equation*}%
leads to (\ref{smik}). Theorem is proved.

Moreover (\ref{smyk}) reduces to (see the second equation in (\ref{smik}))%
\begin{equation*}
\partial _{\tau (\zeta )}e^{\alpha u}=(Q(p(\zeta ))e^{\alpha u})_{x}
\end{equation*}%
It is easy to see the conservation law density $e^{\alpha u}$ is nothing but
a momentum density $A^{0}$ (up to unessential additive and multiplicative
constants).

\subsection{Integrable three dimensional quasilinear equations of the second
order}

Compatibility conditions $(p_{t^{k}})_{t^{n}}=(p_{t^{n}})_{t^{k}}$ lead to
integrable three dimensional hydrodynamic type systems. If $k=1$ and $n=-1$,
then three dimensional hydrodynamic type system%
\begin{equation}
u_{t^{-1}}+v_{t^{0}}+\alpha vu_{t^{0}}=0,\text{ \ \ }v_{t^{1}}=\gamma
vu_{t^{0}}  \label{sis}
\end{equation}%
is determined by the dispersionless Lax pair (see (\ref{neg}) and (\ref{cos}%
))%
\begin{equation}
p_{t^{1}}=\partial _{t^{0}}(u+U(p)),\text{ \ \ }p_{t^{-1}}=\partial
_{t^{0}}\left( \frac{v}{U^{\prime }(p)}\right) ,  \label{lunk}
\end{equation}%
where%
\begin{equation*}
u=\frac{1}{\alpha }\ln (\alpha A^{0}+\sigma ),\text{ \ \ }v=\frac{\gamma }{%
\gamma A^{-1}-\sigma }.
\end{equation*}%
Hydrodynamic type system (\ref{sis}) can be written in the conservative form%
\begin{equation*}
(e^{\alpha u})_{t^{-1}}+\alpha (ve^{\alpha u})_{t^{0}}=0,\text{ \ \ }(\ln
v)_{t^{1}}=\gamma u_{t^{0}}.
\end{equation*}%
Introducing the potential function $z$ such that (see the first conservation
law)%
\begin{equation*}
u=\frac{1}{\alpha }\ln z_{t^{0}},\text{ \ \ \ }v=-\frac{z_{t^{-1}}}{\alpha
z_{t^{0}}},
\end{equation*}%
(\ref{sis}) reduces to the three dimensional quasilinear equation of the
second order%
\begin{equation*}
\alpha (z_{t^{0}}z_{t^{-1}t^{1}}-z_{t^{-1}}z_{t^{0}t^{1}})=\gamma
z_{t^{-1}}z_{t^{0}t^{0}};
\end{equation*}%
introducing the potential function $\tilde{z}$ such that (see the second
conservation law)%
\begin{equation*}
v=e^{\tilde{z}_{t^{0}}},\text{ \ \ }u=\frac{\tilde{z}_{t^{1}}}{\gamma },
\end{equation*}%
(\ref{sis}) reduces to another three dimensional quasilinear equation of the
second order%
\begin{equation}
\tilde{z}_{t^{1}t^{-1}}+e^{\tilde{z}_{t^{0}}}(\gamma \tilde{z}%
_{t^{0}t^{0}}+\alpha \tilde{z}_{t^{0}t^{1}})=0.  \label{old}
\end{equation}%
A substitution (where the potential function $\psi $ is introduced for (\ref%
{lunk}), see (\ref{rez}))%
\begin{equation*}
u=\psi _{t^{1}}+\frac{1}{\alpha }\ln (1-\alpha e^{\gamma \psi _{t^{0}}}),%
\text{ \ \ \ }v=\frac{\gamma e^{\gamma \psi _{t^{0}}}}{1-\alpha e^{\gamma
\psi _{t^{0}}}}\psi _{t^{-1}}
\end{equation*}%
into (\ref{sis}) implies the integrable quasilinear three dimensional
equation of the second order%
\begin{equation*}
\psi _{t^{1}t^{-1}}+\frac{\gamma e^{\gamma \psi _{t^{0}}}}{1-\alpha
e^{\gamma \psi _{t^{0}}}}\psi _{t^{-1}}(\alpha \psi _{t^{0}t^{1}}+\gamma
\psi _{t^{0}t^{0}})=0.
\end{equation*}

If $k=1$ and $n=2$, then the dispersionless Lax pair (see (\ref{gem}) and (%
\ref{agen}))%
\begin{equation}
p_{t^{1}}=\partial _{t^{0}}(u+U(p)),\text{ \ \ }p_{t^{2}}=\partial
_{t^{0}}(w+sU^{\prime }(p))  \label{cunk}
\end{equation}%
determines the constraint $s=e^{\alpha u}$ and the three dimensional
hydrodynamic type system written in the conservative form%
\begin{equation}
\alpha (e^{\alpha u})_{t^{1}}=(\alpha w-\gamma e^{\alpha u})_{t^{0}},\text{
\ \ }u_{t^{2}}=w_{t^{1}}.  \label{syn}
\end{equation}%
Introducing the potential function $z$ such that (see the first conservation
law)%
\begin{equation*}
u=\frac{1}{\alpha }\ln z_{t^{0}},\text{ \ \ }w=z_{t^{1}}+\frac{\gamma }{%
\alpha }z_{t^{0}},
\end{equation*}%
(\ref{syn}) reduces to the three dimensional quasilinear equation of the
second order%
\begin{equation*}
z_{t^{0}t^{2}}=z_{t^{0}}(\alpha z_{t^{1}t^{1}}+\gamma z_{t^{0}t^{1}});
\end{equation*}%
introducing the potential function $\bar{z}$ such that (see the second
conservation law)%
\begin{equation*}
u=\bar{z}_{t^{1}},\text{ \ \ }w=\bar{z}_{t^{2}},
\end{equation*}%
(\ref{syn}) reduces to another three dimensional quasilinear equation of the
second order%
\begin{equation}
\bar{z}_{t^{0}t^{2}}=e^{\alpha \bar{z}_{t^{1}}}(\alpha \bar{z}%
_{t^{1}t^{1}}+\gamma \bar{z}_{t^{0}t^{1}}).  \label{new}
\end{equation}%
A substitution (where the potential function $\psi $ is introduced for (\ref%
{cunk}), see (\ref{rez}))%
\begin{equation*}
u=\psi _{t^{1}}+\frac{1}{\alpha }\ln (1-\alpha e^{\gamma \psi _{t^{0}}}),%
\text{ \ \ }w=\psi _{t^{2}}-\gamma e^{\gamma \psi _{t^{0}}+\alpha \psi
_{t^{1}}}
\end{equation*}%
into (\ref{syn}) implies the integrable quasilinear three dimensional
equation of the second order%
\begin{equation*}
\psi _{t^{0}t^{2}}=e^{\alpha \psi _{t^{1}}}(1-\alpha e^{\gamma \psi
_{t^{0}}})(\gamma \psi _{t^{0}t^{1}}+\alpha \psi _{t^{1}t^{1}}).
\end{equation*}

\textbf{Remark}: The whole hierarchy of positive and negative commuting
flows possesses the symmetry $t^{k+1}\leftrightarrow t^{-k}$. Without loss
of generality it is enough to prove for the independent variables $%
t^{1}\leftrightarrow t^{0}$ only. Indeed, first generating function of
conservation laws (\ref{cos})%
\begin{equation*}
p_{t}=\partial _{x}(u+U(p))
\end{equation*}%
can be rewritten in the same form%
\begin{equation*}
q_{x}=\partial _{t}(v+\tilde{U}(q)),
\end{equation*}%
where $q=U(p),p=\tilde{U}(q),v_{t}=-u_{x}$ (see (\ref{cink})). Since (see (%
\ref{rem}) where $\epsilon =0$ and (\ref{rez})) $U^{\prime \prime
}(p)=\alpha U^{\prime ^{2}}(p)+\gamma U^{\prime }(p)$, then $\tilde{U}%
^{\prime \prime }(q)=-\gamma \tilde{U}^{\prime ^{2}}(q)-\alpha \tilde{U}%
^{\prime }(q)$. Thus, equations (\ref{old}) and (\ref{new}) are equivalent
to each other (by an appropriate change of the constants $\alpha
\leftrightarrow \gamma $ and the independent variables $t^{1}\leftrightarrow
t^{0},t^{2}\leftrightarrow t^{-1}$).

Let us consider the three dimensional hydrodynamic type system written in
the conservative form (see (\ref{agen}), (\ref{smyk}) and (\ref{rez})) 
\begin{equation}
\partial _{\tau }e^{\alpha u}=\partial _{t^{0}}\frac{\gamma e^{\alpha u}}{%
1-\alpha e^{\gamma p}},\ \ \partial _{t^{1}}p=\partial _{t^{0}}\left( u-%
\frac{1}{\alpha }\ln (1-\alpha e^{\gamma p})\right) .  \label{last}
\end{equation}%
(\ref{last}) reduces to the integrable three dimensional quasilinear
equation of the second order%
\begin{equation*}
\alpha (z_{t^{0}}z_{\tau t^{1}}-z_{\tau }z_{t^{0}t^{1}})=\left( z_{\tau
}-\gamma z_{t^{0}}\right) z_{\tau t^{0}},
\end{equation*}%
where the function $z$ is a potential of the first conservation law; (\ref%
{last}) reduces to another integrable three dimensional quasilinear equation
of the second order%
\begin{equation*}
\psi _{\tau t^{1}}-\gamma \psi _{t^{0}t^{1}}=e^{\gamma \psi _{t^{0}}}(\gamma
\psi _{\tau t^{0}}+\alpha \psi _{\tau t^{1}}),
\end{equation*}%
where the function $\psi $ is a potential of the second conservation law.

\section{Conclusion}

In this paper, integrable Hamiltonian hydrodynamic chains associated with
Dorfman Poisson brackets are described. These Dorfman Poisson brackets are
parameterized by six ($\alpha ,\beta ,\gamma ,\delta ,\epsilon ,\xi $)
constants. However, three distinguish cases are selected by special values
of two constants $\beta $ and $\delta $ only. The first and most general
class is determined by arbitrary parameters ($\alpha \neq 0,\beta \neq
0,\gamma \neq 0,\delta \neq 0$); the second class is determined by the
restriction $\beta =0$ while $\delta \neq 0$; the third class is determined
by more deep restriction $\beta =0$ and $\delta =0$. In the particular case $%
\delta =0$, hydrodynamic chains of the first class were constructed in \cite%
{algebra}; corresponding Hamiltonian reductions are the Hamiltonian
chromatography systems. Hydrodynamic chains of the second class also are
well known. In a particular case (a special limit $\alpha =0$), they are
nothing but the Kupershmidt hydrodynamic chains (see \cite{Blaszak}, \cite%
{KuperNorm}, \cite{Manas} and \cite{MaksKuper}). The most investigated
Benney hydrodynamic chain (see \cite{Benney}, \cite{Gibbons}, \cite{GR}, 
\cite{GT}, \cite{Gib+Yu}, \cite{kod+water}, \cite{KM}, \cite{zakh}) belongs
to the third class.

In a general case, hydrodynamic chains of the third class are considered in 
\cite{MaksGen}. However, the simplest hydrodynamic chain (\ref{sem}) was
missed. This hydrodynamic chain is found (as well as corresponding
hydrodynamic chain (\ref{pyat}) of the second class) in the presented paper.
Corresponding Hamiltonian density depends on \textit{infinitely many} field
variables (moments $A^{k}$). This phenomenon was never mentioned in a
literature.

All integrable hydrodynamic chains possess infinite series of conservation
laws and commuting flows. Corresponding generating functions are
constructed. These hydrodynamic chains are extended on negative values of
moments $A^{k}$ as well as associated integrable hierarchies are extended on
negative values of time variables $t^{n}$.

Most obvious and simple three dimensional hydrodynamic type systems are
presented. They are converted to the three dimensional quasilinear equations
of the second order (see a general theory in \cite{BFT}). Most of them are
new. Nevertheless, the list of these equations (as well as associated
hydrodynamic type systems) is so large, that all already found examples in 
\cite{FMS} can be extracted from this list by virtue of different parametric
reductions. Moreover, the list presented in \cite{FMS} can be split on three
parts according given approach here. Examples (\textbf{18}) and (\textbf{21}%
) belonging to the Benney hierarchy are of the third class; examples (%
\textbf{20})$_{\text{1}}$, (\textbf{20})$_{\text{2}}$ and (\textbf{22})
belong to the Kupershmidt hierarchy (see \cite{MaksKuper}, where $\beta =3$%
); most complicated examples (\textbf{16}), (\textbf{17}) and (\textbf{19})
belong to the first class, where $V(p)=\wp (p)$ and the elliptic
Weiershtrass function satisfies $\wp ^{\prime ^{2}}(p)=4\wp ^{3}(p)+\bar{%
\delta}^{2}$, while the function $V(p)$ satisfies $2V(p)V^{\prime \prime
}(p)=3(V^{\prime ^{2}}(p)-\bar{\delta}^{2})$ (see (\ref{eq})). This is a
full list of integrable three dimensional Hamiltonian hydrodynamic type
systems (see detail in \cite{FMS})%
\begin{equation*}
\left( 
\begin{array}{c}
u \\ 
v%
\end{array}%
\right) _{t}=\left( 
\begin{array}{cc}
\partial _{x} & 0 \\ 
0 & \partial _{y}%
\end{array}%
\right) \left( 
\begin{array}{c}
h_{u} \\ 
h_{v}%
\end{array}%
\right) .
\end{equation*}%
In another paper \cite{FOS}, a full list of integrable three dimensional
Hamiltonian hydrodynamic type systems%
\begin{equation*}
\left( 
\begin{array}{c}
u \\ 
v%
\end{array}%
\right) _{t}=\left( 
\begin{array}{cc}
0 & \partial _{x} \\ 
\partial _{x} & \partial _{y}%
\end{array}%
\right) \left( 
\begin{array}{c}
h_{u} \\ 
h_{v}%
\end{array}%
\right) 
\end{equation*}%
also is presented. Examples (\textbf{21}) and (\textbf{24}) belong again to
the Benney hierarchy; examples (\textbf{20}), (\textbf{22})$_{\text{1}}$ and
(\textbf{23}) belong to the Kupershmidt hierarchy (see \cite{MaksKuper},
where $\beta =3$); example (\textbf{22})$_{\text{2}}$ belongs to the first
class, where $V(p)=\wp (p)$. And just the last example (\textbf{19}) is not
yet recognized. Thus, most these examples associated with two dimensional
constant Poisson brackets simultaneously are connected with hydrodynamic
chains associated with Dorfman Poisson brackets (but not vice versa!).
However, the question \textquotedblleft how to connect both Hamiltonian
structures\textquotedblright\ is open at this moment.

The relationship between three dimensional hydrodynamic type systems, three
dimensional quasilinear equations of the second order and hydrodynamic
chains described above is very important. All hydrodynamic chains (\ref{lin}%
) presented in this paper are very convenient for more deep investigation by
the moment decomposition approach (see detail in \cite{MaksHam} and in \cite%
{MaksZyk}). This approach allows to extract multi-parametric solutions (see
detail in \cite{maksbenney}) of these hydrodynamic chains and corresponding
three dimensional hydrodynamic type systems as well as three dimensional
quasilinear equations of the second order. Thus, any aforementioned example
can be equipped by a corresponding hydrodynamic chain described in this
paper. It means, that such a hydrodynamic chain can be considered as an
infinite set of the so called pseudo-nonlocalities (i.e. moments) allowing
to extend integrable three dimensional two component hydrodynamic type
systems on infinitely many field variables. It means that a complexity of
integrable three dimensional quasilinear equations of the second order can
be translated to a complexity of two dimensional hydrodynamic chains (i.e.
hydrodynamic type systems containing infinitely many equations).

\section*{Acknowledgements}

I thank Eugene Ferapontov for stimulating and clarifying discussions.

I am grateful to the SISSA in Trieste (Italy) where some part of this work
has been done. This research was particularly supported by the RFBR grant
08-01-00464-a and by the grant of Presidium of RAS \textquotedblleft
Fundamental Problems of Nonlinear Dynamics\textquotedblright .


\addcontentsline{toc}{section}{References}

\begin{thebibliography}{99}
\bibitem{Benney} \emph{D.J. Benney,} \newblock Some properties of long
non-linear waves, Stud. Appl. Math., \textbf{52} (1973) 45-50.

\bibitem{bogdan} \emph{L.V. Bogdanov, B.G. Konopelchenko}, \newblock %
Symmetry constraints for dispersionless integrable equations and systems of
hydrodynamic type, Phys. Lett. A, \textbf{330 }(2004) 448--459.

\bibitem{Blaszak} \emph{M. B\l aszak, B.M. Szablikowski}, \newblock %
Classical $R-$matrix theory of dispersionless systems: I. (1+1)-dimension
theory, J. Phys. A: Math. Gen. \textbf{35} (2002) 10325-10344. \emph{M. B\l %
aszak, B.M. Szablikowski}, \newblock Classical $R-$matrix theory of
dispersionless systems: II. (2+1)-dimension theory, J. Phys. A: Math. Gen. 
\textbf{35} (2002) 10345-10364.

\bibitem{BFT} \emph{P.A. Burovski, E.V. Ferapontov, S.P. Tsarev}, \newblock %
Second order quasilinear PDEs and conformal structures in projective space,
International J. Math. \textbf{21} No. 6 (2010) 799-841.

\bibitem{Dorfman} \emph{I.Ya. Dorfman}, \newblock Dirac structures and
integrability of nonlinear evolution equations; Nonlinear Science: Theory
and Applications, John Wiley \& Sons, New York (1993) 176 pp.

\bibitem{dn} \emph{B.A. Dubrovin, S.P. Novikov,} \newblock Hamiltonian
formalism of one-dimensional systems of hydrodynamic type and the
Bogolyubov-Whitham averaging method, Soviet Math. Dokl., \textbf{27} (1983)
665--669. \emph{B.A. Dubrovin, S.P. Novikov,} \newblock Hydrodynamics of
weakly deformed soliton lattices. Differential geometry and Hamiltonian
theory, Russian Math. Surveys, \textbf{44} No. 6 (1989) 35--124.

\bibitem{FerKar} \emph{E.V. Ferapontov, K.R. Khusnutdinova}, \newblock On
integrability of (2+1)-dimensional quasilinear systems, Comm. Math. Phys., 
\textbf{248} (2004) 187-206, \emph{E.V. Ferapontov, K.R. Khusnutdinova}, %
\newblock The characterization of 2-component (2+1)-dimensional integrable
systems of hydrodynamic type, J. Phys. A: Math. Gen., \textbf{37} No. 8
(2004) 2949 - 2963.

\bibitem{FKMP} \emph{E.V. Ferapontov, K.R. Khusnutdinova, D.G. Marshall,
M.V. Pavlov}, \newblock Classification of Integrable Hydrodynamic chains
associated with Kupershmidt's brackets. J. Maths. Phys., \textbf{47} (2006)
103507-103520.

\bibitem{FerMarsh} \emph{E.V. Ferapontov, D.G. Marshall}, \newblock %
Differential-geometric approach to the integrability of hydrodynamic chains:
the Haantjes tensor, Mathematische Annalen, \textbf{339} No. 1 (2007) 61-99.

\bibitem{FMS} \emph{E.V. Ferapontov, A. Moro, V. V. Sokolov}, \newblock %
Hamiltonian systems of hydrodynamic type in 2+1 dimensions, Comm. Math.
Phys. \textbf{285} No. 1 (2009) 31-65.

\bibitem{FOS} \emph{E.V. Ferapontov, A.V. Odesskii, N.M. Stoilov}, \newblock %
Classification of integrable two component Hamiltonian systems of
hydrodynamic type in 2+1 dimensions, submitted (2010) arXiv:1007.3782.

\bibitem{Gibbons} \emph{J. Gibbons,} \newblock Collisionless Boltzmann
equations and integrable moment equations, Physica D, \textbf{3} (1981)
503-511.

\bibitem{GR} \emph{J. Gibbons, A. Raimondo}, \newblock Differential geometry
of hydrodynamic Vlasov equations. J. Geom. and Phys. \textbf{57} (2007)
1815-1828.

\bibitem{GT} \emph{J. Gibbons, S.P. Tsarev}, \newblock Reductions of the
Benney equations, Phys. Lett. A, \textbf{211} (1996) 19-24. \emph{J.
Gibbons, S.P. Tsarev}, \newblock Conformal maps and reductions of the Benney
equations, Phys. Lett. A, \textbf{258} (1999) 263-271.

\bibitem{Gib+Yu} \emph{J. Gibbons, Yu. Kodama,} \newblock Integrable
quasilinear systems: generalized hodograph transformation. Nonlinear
evolutions (Balaruc-les-Bains, 1987), 97--107, World Sci. Publ., Teaneck,
NJ, 1988. \emph{Yu. Kodama,} \emph{J. Gibbons, }\newblock A method for
solving the dispersionless KP hierarchy and its exact solutions. II. Phys.
Lett. A 135 (1989), No. 3, 167--170. \emph{Yu. Kodama, J. Gibbons,} 
\newblock
Integrability of the dispersionless KP hierarchy. Nonlinear world, Vol. 1
(Kiev, 1989), 166--180, World Sci. Publ., River Edge, NJ, 1990. \emph{J.
Gibbons, L.A. Yu}, \newblock The initial value problem for reductions of the
Benney equations, Inverse Problems \textbf{16} No. 3 (2000) 605-618, \emph{%
L.A. Yu}, \newblock Waterbag reductions of the dispersionless discrete KP
hierarchy, J. Phys. A: Math. Gen., \textbf{33} (2000) 8127--8138.

\bibitem{kod+water} \emph{Yu. Kodama}, \newblock A method for solving the
dispersionless KP equation and its exact solutions. Phys. Lett. A, \textbf{%
129} No. 4 (1988) 223-226. \emph{Yu. Kodama}, \newblock A solution method
for the dispersionless KP equation, Prog. Theor. Phys. Supplement. \textbf{94%
} (1988) 184. \emph{Yu. Kodama}, \newblock Solutions of the dispersionless
Toda equation, Phys. Lett. A, \textbf{147} No. 8-9 (1990) 477-482. \emph{Yu.
Kodama}, \newblock Exact solutions of hydrodynamic type equations having
infinitely many conserved densities, Phys. Lett. A, \textbf{135} No. 3
(1989) 171-174.

\bibitem{KuperNorm} \emph{B.A. Kupershmidt}, \newblock Deformations of
integrable systems, Proc. Roy. Irish Acad. Sect. A, \textbf{83} No. 1 (1983)
45-74. \emph{B.A. Kupershmidt}, \newblock Normal and universal forms in
integrable hydrodynamical systems, Proceedings of the Berkeley-Ames
conference on nonlinear problems in control and fluid dynamics (Berkeley,
Calif., 1983), in Lie Groups: Hist., Frontiers and Appl. Ser. B: Systems
Inform. Control, II, Math Sci Press, Brookline, MA, (1984) 357-378.

\bibitem{KuperPavl} \emph{B.A. Kupershmidt}, \newblock Hydrodynamic chains
of Pavlov class, Phys. Lett. A, \textbf{356} (2006) 115-118.

\bibitem{KM} \emph{B.A. Kupershmidt, Yu.I. Manin}, \newblock Long wave
equations with a free surface. I. Conservation laws and solutions. (Russian)
Func. Anal. Appl. \textbf{11} No. 3 (1977) 31--42. \emph{B.A. Kupershmidt,
Yu.I. Manin}, \newblock Long wave equations with a free surface. II. The
Hamiltonian structure and the higher equations. (Russian) Func. Anal. Appl. 
\textbf{12} No. 1 (1978) 25--37. \emph{D.R. Lebedev, Yu.I. Manin}, \newblock %
Conservation laws and representation of Benney's long wave equations, Phys.
Lett. A, \textbf{74} No. 3,4 (1979) 154-156.

\bibitem{ls} \emph{M.A. Lavrentiev, B.V. Shabat}, 
\newblock {Metody teorii
funktsi\u\i kompleksnogo peremennogo}(Russian) [Methods of the theory of
functions of a complex variable] Third corrected edition Izdat.
\textquotedblleft Nauka\textquotedblright , Moscow (1965) 716 pp. \emph{P.
Henrici}, \newblock {Metody teorii
funktsi\u\i kompleksnogo peremennogo}Topics in computational complex
analysis. IV. The Lagrange-B\"{u}rmann formula for systems of formal power
series. Computational aspects of complex analysis (Braunlage, 1982),
193--215, NATO Adv. Sci. Inst. Ser. C, Math. Phys. Sci., 102, Reidel,
Dordrecht, 1983.

\bibitem{Manas} \emph{M. Manas}, \newblock$S-$functions, reductions and
hodograph solutions of the $r$th dispersionless modified KP and Dym
hierarchies, J. Phys. A: Math. Gen., \textbf{37} (2004) 11191--11221.

\bibitem{OPS} \emph{A.V. Odesski, M.V. Pavlov, V.V. Sokolov}, \newblock %
Classification of integrable Vlasov-type equations, Theor. and Math. Phys., 
\textbf{154} No. 2 (2008) 209--219.

\bibitem{MaksZiemek} \emph{M. V. Pavlov, Z. Popowicz,} \newblock A complete
classification of integrable generalized Hamiltonian two component 2+1
hydrodynamic type systems.

\bibitem{MaksZyk} \emph{M.V. Pavlov, S.A. Zykov}, \newblock Classification
of integrable conservative hydrodynamic chains, submitted (2010) arXiv:
0912.4954.

\bibitem{MaksEps} \emph{M.V. Pavlov}, \newblock Integrable hydrodynamic
chains, J. Math. Phys., \textbf{44} No. 9 (2003) 4134-4156.

\bibitem{MaksKuper} \emph{M.V. Pavlov}, \newblock The Kupershmidt
hydrodynamic chains and lattices, IMRN (2006) article ID 46987.

\bibitem{algebra} \emph{M.V. Pavlov}, \newblock Algebro-geometric approach
in the theory of integrable hydrodynamic type systems, Comm. Math. Phys., 
\textbf{272} No. 2 (2007) 469-505.

\bibitem{MaksGen} \emph{M.V. Pavlov}, \newblock Classification of integrable
hydrodynamic chains and generating functions of conservation laws, J. Phys.
A: Math. Gen., (2006) 10803-10819.

\bibitem{MaksHam} \emph{M.V. Pavlov}, \newblock The Hamiltonian approach in
the classification and the integrability of hydrodynamic chains, ArXiv:
Nlin.SI/0603057.

\bibitem{maksbenney} \emph{M.V. Pavlov}, \newblock Integrability of the
Gibbons---Tsarev system, Amer. Math. Soc. Transl., (2) \textbf{224} (2008)
247-259.

\bibitem{Tsar} \emph{S.P. Tsarev}, \newblock On Poisson brackets and
one-dimensional Hamiltonian systems of hydrodynamic type, Soviet Math.
Dokl., \textbf{31} (1985) 488--491. \emph{S.P. Tsarev}, \newblock The
geometry of Hamiltonian systems of hydrodynamic type. The generalized
hodograph method, Math. USSR Izvestiya, \textbf{37} No. 2 (1991) 397--419.

\bibitem{zakh} \emph{V.E. Zakharov}, \newblock Benney's equations and
quasi-classical approximation in the inverse problem method, Funct. Anal.
Appl., \textbf{14} No. 2 (1980) 89-98. \emph{V.E. Zakharov}, \newblock On
the Benney's Equations, Physica \textbf{3}D (1981) 193-200.
\end{thebibliography}
\end{document}